\documentclass[
prd,preprint,aps,endfloats,tightenlines,
showpacs,superscriptaddress,nofootinbib,
amssymb
]{revtex4}
\usepackage{graphicx}
\begin{document}
\setlength{\baselineskip}{16pt}
%
%
\title{
Calculation of $K \to \pi\pi$ decay amplitudes
with improved Wilson fermion action in lattice QCD
}
%
%
\author{ N.~Ishizuka }
\affiliation{
Center for Computational Sciences,
University of Tsukuba,
Tsukuba, Ibaraki 305-8577, Japan
}
\affiliation{
Graduate School of Pure and Applied Sciences,
University of Tsukuba,
Tsukuba, Ibaraki 305-8571, Japan
}
%
\author{ K.-I.~Ishikawa }
\affiliation{
Department of Physics,
Hiroshima University,
Higashi-Hiroshima, Hiroshima 739-8526, Japan
}
%
\author{ A.~Ukawa }
\affiliation{
RIKEN Advanced Institute for Computational Science,
Kobe 650-0047, Japan
}
%
\author{ T.~Yoshi\'{e} }
\affiliation{
Center for Computational Sciences,
University of Tsukuba,
Tsukuba, Ibaraki 305-8577, Japan
}
\affiliation{
Graduate School of Pure and Applied Sciences,
University of Tsukuba,
Tsukuba, Ibaraki 305-8571, Japan
}
%
%
\date{ \today }
%
%
\begin{abstract}
We present our result for
the $K\to\pi\pi$ decay amplitudes
for both the $\Delta I=1/2$ and $3/2$ processes
with the improved Wilson fermion action.
Expanding on the earlier works
by Bernard {\it et al.} and by Donini {\it et al.},
we show that mixings with four-fermion operators with wrong chirality
are absent even for the Wilson fermion action
for the parity odd process in both channels
due to CPS symmetry.
Therefore, after subtraction of an effect from the lower dimensional operator,
a calculation of the decay amplitudes is possible
without complications from operators with wrong chirality,
as for the case with chirally symmetric lattice actions.
As a first step to verify the possibility of calculations
with the Wilson fermion action,
we consider the decay amplitudes at an unphysical quark mass $m_K \sim 2 m_\pi$.
Our calculations are carried out
with $N_f=2+1$ gauge configurations
generated with the Iwasaki gauge action and
nonperturbatively $O(a)$-improved Wilson fermion action
at $a=0.091\,{\rm fm}$,
$m_\pi=280\,{\rm MeV}$, and $m_K=580\,{\rm MeV}$
on a $32^3\times 64$ ($La=2.9\,{\rm fm}$) lattice.
For the quark loops in the penguin and disconnected contributions
in the $I=0$ channel,
the combined hopping parameter expansion and truncated solver method
work very well for variance reduction.
We obtain, for the first time with a Wilson-type fermion action, that
${\rm Re}A_0 = 60(36) \times10^{ -8}\,{\rm GeV}$ and
${\rm Im}A_0 =-67(56) \times10^{-12}\,{\rm GeV}$
for a matching scale $q^* =1/a$.
The dependence on the matching scale $q^*$ for these values is weak.
\end{abstract}
\pacs{ 11.15.Ha, 12.38.Gc, 13.25.Es }
\maketitle
%
%
\section{ Introduction }
\label{Sec:Introduction}
Calculation of the $K\to\pi\pi$ decay amplitudes
is very important to quantitatively understand the $\Delta I=1/2$ rule
in the decay of the neutral $K$ meson system and
to theoretically predict the direct $CP$ violation parameter
($\epsilon'/\epsilon$)
from the Standard Model.
A direct lattice QCD calculation of the decay amplitudes
for the $\Delta I=3/2$ process has been attempted for a long time.
Recently,
the RBC-UKQCD Collaboration presented the results
at the physical quark mass in Ref.~\cite{A2:RBC-UKQCD_phys},
and those in the continuum limit in Ref.~\cite{A2:RBC-UKQCD_phys_cont}
in the physical kinematics,
where the pions in the final state have finite momenta.
They used the domain wall fermion action
which preserves chiral symmetry on the lattice.

A direct calculation of the decay amplitudes
for the $\Delta I=1/2$ process has been unsuccessful for a long time,
due to large statistical fluctuations from the disconnected diagrams.
A first direct calculation
was reported by the RBC-UKQCD Collaboration
in Ref.~\cite{A0:RBC-UKQCD_400} at a lattice spacing $a=0.114\,{\rm fm}$
and a pion mass $m_\pi=422\,{\rm MeV}$
on a $16^3\times 32$ lattice with the domain wall fermion action.
They also presented a result at a smaller quark mass
($m_\pi=330\,{\rm MeV}$) on a $24^3\times 64$ lattice
with the same fermion action at Lattice 2011~\cite{A0:RBC-UKQCD_300}.
In these two calculations,
the kinematics was a $K$ meson at rest decaying to two zero momentum pions
at an unphysical quark mass satisfying $m_K \sim 2 m_\pi$.
The RBC-UKQCD Collaboration has since been attempting
a direct calculation in the physical kinematics
at the physical quark mass by utilizing $G$-parity boundary conditions.
Their preliminary result was reported
at Lattice 2014~\cite{A0:RBC-UKQCD_phys}.

An aim of the present article is to report on our calculation
of the $K\to\pi\pi$ decay amplitudes with the improved Wilson fermion action
for both the $\Delta I=1/2$ and $3/2$ processes.
That such a calculation is feasible stems from a realization,
as shown in the present article,
that CPS symmetry~\cite{CPS:Bernard} and its extensions~\cite{CPS:Donini}
ensure that
mixings with four-fermion operators with wrong chirality
are absent even for the Wilson fermion action
for the parity odd process in both channels.
A mixing to a lower dimension operator does occur,
which gives unphysical contributions to the amplitudes on the lattice.
However, it can be nonperturbatively subtracted
by imposing a renormalization condition~\cite{sub:Dawson,sub:Testa}.
After the subtraction we can obtain the physical decay amplitudes
by the renormalization factor having the same structure
as for the chiral symmetry preserved case.
A potential advantage with the Wilson fermion action
over chirally symmetric lattice actions such as the domain wall action
is that the computational cost is generally smaller.
Hence, with the same amount of computational resources,
a statistical improvement may be expected with the lattice calculation
of the decay amplitudes, albeit this point
has to be verified by actual calculations.

In the present work,
we consider the decay of $K$ meson to two zero momentum pions
at an unphysical quark mass $m_K \sim 2 m_\pi$,
as in Refs.~\cite{A0:RBC-UKQCD_400,A0:RBC-UKQCD_300},
as the first step of a study with the Wilson fermion action.
Our calculations are carried out
on a subset of gauge configurations previously generated
by the PACS-CS Collaboration
with the Iwasaki gauge action
and the nonperturbatively $O(a)$-improved Wilson fermion action
for $N_f=2+1$ flavors
at $\beta=1.9$ on a $32^3\times 64$ lattice~\cite{conf:PACS-CS}.
The subset corresponds to the hopping parameters
$\kappa_{ud}=0.13770$ for the up and the down quark and
$\kappa_{s }=0.13640$ for the strange quark.
We further generate gauge configurations
at the same parameters to improve the statistics.
The total number of gauge configurations used in the present work is $480$.
The parameters determined from the hadron spectrum analysis
are $a=0.091\ {\rm fm}$ for the lattice spacing,
$La=2.91\ {\rm fm}$ for the lattice size,
$m_\pi=275.7(1.5)\,{\rm MeV}$ and $m_K=579.7(1.3)\,{\rm MeV}$
for the pion and the $K$ masses.
The energy of the two-pion state is shifted from $2 m_\pi$
by the two-pion interaction on the lattice.
The energy difference between the initial $K$ meson
and the final two-pion state takes a nonzero value,
$\Delta E = 21(3)\,{\rm MeV}$ for the $I=2$ channel,
and $36(18)\,{\rm MeV}$ for the $I=0$ channel on these configurations.
In the present work we assume that these mismatches of the energy
give only small effects to the decay amplitudes.

This paper is organized as follows.
The $K\to\pi\pi$ decay amplitudes can be calculated
from the product of the $K\to\pi\pi$ matrix elements
of the $\Delta S=1$ four-fermion weak interaction operators
and the Wilson coefficient functions for the operator product expansion.
In Sec.~\ref{Sec: Delta S=1 operators}
these four-fermion operators are introduced and
the operator mixing among them for the Wilson fermion action is
discussed.
In Sec.~\ref{Sec: Method}
we describe the method of calculation used in the present work.
The simulation parameters are also given.
We present our results in Sec.~\ref{Sec: Results},
and compare them with those by the RBC-UKQCD Collaboration
and the experimental values.
Conclusions of the present work are given
in Sec.~\ref{Sec:Conclusions}.

Our calculations have been carried out
on the PACS-CS computer and T2K-Tsukuba at University of Tsukuba,
the K computer at the RIKEN Advanced Institute for Computational Science,
SR16000 at University of Tokyo, and
SR16000 at High energy Accelerator Research Organization (KEK).
Our preliminary results have been reported
at Lattice 2013 and 2014~\cite{KPIPI:Our_old}.
%
%
\section{ $\Delta S=1$ operators }
\label{Sec: Delta S=1 operators}
%
\subsection{ $\Delta S=1$ weak operators in the continuum theory }
\label{sec:continuum}
The effective Hamiltonian of the $K\to\pi\pi$ decay in the continuum theory
can be written as~\cite{Review:Buras}
\begin{equation}
  H = \frac{ G_F }{\sqrt{2}} \left( V_{us}^* V_{ud} \right)
      \sum_{i=1}^{10} \left( z_{i}(\mu) + \tau y_{i}(\mu) \right) Q_{i}(\mu)
\ ,
\label{eq:weak_Hamlitonian}
\end{equation}
with $\tau = - \left( V_{ts}^* V_{td} \right) / \left( V_{us}^* V_{ud} \right)$, and
$z_{i}(\mu)$ and $y_{i}(\mu)$ ($i=1,2,\cdots 10$)
are the coefficient functions at renormalization scale $\mu$.
Here we consider the case $\mu \le m_{c}$,
where three light quarks, up, down and strange,
are the active quarks in the theory.
The ten operators $Q_{i}(\mu)$ ($i=1,2,\cdots 10$)
denote the $\Delta S=1$ four-fermion operators
renormalized at $\mu$, which are given by
\begin{eqnarray}
&&
Q_{1 }=(\bar{s}       d)(\bar{u}       u)_{LL}
\ ,
\label{eq:weak_op1}
\\
&&
Q_{2 }=(\bar{s}\times d)(\bar{u}\times u)_{LL}
\ ,
\label{eq:weak_op2}
\\
&&
Q_{3 }=(\bar{s}       d) \sum_{q} (\bar{q}       q)_{LL}
\ ,
\label{eq:weak_op3}
\\
&&
Q_{4 }=(\bar{s}\times d) \sum_{q} (\bar{q}\times q)_{LL}
\ ,
\label{eq:weak_op4}
\\
&&
Q_{5 }=(\bar{s}       d) \sum_{q} (\bar{q}       q)_{LR}
\ ,
\label{eq:weak_op5}
\\
&&
Q_{6 }=(\bar{s}\times d) \sum_{q} (\bar{q}\times q)_{LR}
\ ,
\label{eq:weak_op6}
\\
&&
Q_{7 }=\frac{3}{2}\, (\bar{s}       d) \sum_{q} e_q (\bar{q}       q)_{LR}
\ ,
\label{eq:weak_op7}
\\
&&
Q_{8 }=\frac{3}{2}\, (\bar{s}\times d) \sum_{q} e_q (\bar{q}\times q)_{LR}
\ ,
\label{eq:weak_op8}
\\
&&
Q_{9 }=\frac{3}{2}\, (\bar{s}       d) \sum_{q} e_q (\bar{q}       q)_{LL}
\ ,
\label{eq:weak_op9}
\\
&&
Q_{10}=\frac{3}{2}\, (\bar{s}\times d) \sum_{q} e_q (\bar{q}\times q)_{LL}
\ ,
\label{eq:weak_op10}
\end{eqnarray}
where $(\bar{s}d)(\bar{u}u)_{L,R/L}
= ( \bar{s} \gamma_\mu ( 1 - \gamma_5 ) d )( \bar{u} \gamma_\mu ( 1 \pm \gamma_5 ) u)$,
and $\times$ means the contraction of color indices according to
$
    (\bar{s}\times d)_L (\bar{u}\times d)_L
  = \sum_{a,b} (\bar{s}_a   d_b)_L (\bar{u}_b d_a)_L
$.
The summation for $q$ is taken for the active quarks ($q=u,d,s$)
and the electric charge takes values $e_{u}=+2/3$ and $e_{d}=e_{s}=-1/3$.

In the four-dimensional space-time
these operators are not all independent, satisfying the relations :
\begin{eqnarray}
&&  Q_{4 } = - Q_{1} + Q_{2} + Q_{3}
\ ,
\label{eq:op_rel1}
\\
&&
  Q_{9 } = ( 3 Q_{1} - Q_{3} )/2
\ ,
\label{eq:op_rel2}
\\
&&
  Q_{10} = ( 3 Q_{2} - Q_{4} )/2 = Q_{2} + ( Q_{1} - Q_{3} )/2
\ ,
\label{eq:op_rel3}
\end{eqnarray}
due to the Fierz identity.
In general dimensions, however, these relations are not valid.
Therefore, if we adopt the dimensional regularization for regularization,
we should deal with all operators $Q_{i}$ for $i=1,2,\cdots 10$
as independent.
%
\subsection{ Operator mixing for the Wilson fermion action }
The matrix elements calculated on the lattice are converted to those in the continuum
by the renormalization factor for the operators.
In this section we discuss the renormalization factor
in the case of the Wilson fermion action.

As already mentioned in Sec.~\ref{sec:continuum},
the 10 four-fermion operators $Q_i$ are not independent,
and they may be arranged into 7 linearly independent combinations
according to the irreducible representation
of the flavor ${\rm SU}(3)_L \times {\rm SU}(3)_R$ symmetry group.
The 7 operators consist of
$({\bf 27},{\bf 1}) + 4\cdot ({\bf 8},{\bf 1}) + 2\cdot ({\bf 8},{\bf 8})$,
whose components are given by
\begin{eqnarray}
({\bf 27},{\bf 1}) &&  Q_{1}' =   3 Q_{1} + 2 Q_{2} - Q_{3}
\ ,
\label{eq:chiral_rep1}
\\
({\bf 8},{\bf 1})  &&  Q_{2}' =   2 Q_{1} - 2 Q_{2} + Q_{3} \ , \cr
                   &&  Q_{3}' = - 3 Q_{1} + 3 Q_{2} + Q_{3} \ , \cr
                   &&  Q_{5} \ , \quad Q_{6}
\ ,
\label{eq:chiral_rep2}
\\
({\bf 8},{\bf 8}) &&  Q_7 \ , \quad Q_8
\ .
\label{eq:chiral_rep3}
\end{eqnarray}
The operators $Q_{1,2,3}'$ are the $LL$ type four-fermion operators and
$Q_{5,6,7,8}$ are of $LR$ type.

If the chiral symmetry is preserved in the regularization,
mixings between operators in different representations are forbidden.
For the Wilson fermion action, however,
chiral symmetry is broken to the vector subgroup,
${\rm SU}(3)_L \times {\rm SU}(3)_R \to {\rm SU}(3)_V$.
Hence mixings among different representations is in general allowed,
and new operators arise through radiative corrections.
However, we show below that such a problem is absent for the parity odd part
of the operators in the list of (\ref{eq:weak_op1})--(\ref{eq:weak_op10}) or
of (\ref{eq:chiral_rep1})--(\ref{eq:chiral_rep3})
for the Wilson fermion action employed in the present work.

To investigate the operator mixing,
we exploit the full set of unbroken symmetries for the Wilson fermion action,
namely flavor ${\rm SU}(3)_V$, parity $P$, charge conjugation $C$,
and $CPS$ which is the symmetry under $CP$ transformation
followed by the exchange of the $d$ and the $s$ quarks.
All operators in the list
(\ref{eq:chiral_rep1})--(\ref{eq:chiral_rep3})
are $CPS=+1$ operators,
but the following operators also have the same quantum numbers including $CPS$,
\begin{eqnarray}
&&
Q_X = (\bar{s}       d) (\bar{d}       d - \bar{s}       s )_{SP+PS}
\ ,
\label{eq:op_2a}
\\
&&
Q_Y = (\bar{s}\times d) (\bar{d}\times d - \bar{s}\times s )_{SP+PS}
\ ,
\label{eq:op_2b}
\end{eqnarray}
where
$ (\bar{s}d)(\bar{d}d )_{SP+PS}
= (\bar{s}d)_S (\bar{d}d )_P + (\bar{s}d)_P (\bar{d}d )_S$ and
$(\bar{s}d)_S = \bar{s} d$ and
$(\bar{s}d)_P = \bar{s} \gamma_5 d$.
Therefore
we have to consider mixings including these operators.

Let us discuss the problem in two steps,
first considering mixings through diagrams in which gluons are exchanged
between quarks of the four-fermion operators (gluon exchange diagrams),
and second through penguin diagrams in which a pair of quarks
from the four-fermion operators forms a quark loop.

For the first type of mixings,
it was shown in Ref.~\cite{CPS:Donini} that
the parity odd part of the $LL$ and $LR$ type operators,
and the $SP+PS$ type operator do not mix with each other.
One can prove this through the use of $CPS$, $CPS'$ and $CPS''$ symmetries
which holds for the gluon exchange diagrams,
where $S'$ and $S''$ are the flavor switching
for a four-fermion operator
$(\bar{\psi}_1        \psi_2)(\bar{\psi_3}        \psi_4)_{\Gamma_1\Gamma_2}$ or
$(\bar{\psi}_1 \times \psi_2)(\bar{\psi_3} \times \psi_4)_{\Gamma_1\Gamma_2}$
defined by
\begin{eqnarray}
&&
  S'  \quad : \quad \psi_1 \leftrightarrow \psi_2 \ ,
              \quad \psi_3 \leftrightarrow \psi_4
\ , \\
&&
 S'' \quad : \quad \psi_1 \leftrightarrow \psi_4 \ ,
             \quad \psi_2 \leftrightarrow \psi_3
\ .
\end{eqnarray}
The parity odd part of the $LL$ and $LR$ type operators
in (\ref{eq:chiral_rep1})--(\ref{eq:chiral_rep3}),
which are of $-VA-AV$ and $VA-AV$ type,
and that of $Q_{X,Y}$ in (\ref{eq:op_2a})--(\ref{eq:op_2b}),
which is of $SP+PS$ type,
are eigenvectors of the $CPS'$ and $CPS''$ symmetry
with a different set of eigenvalues,
\begin{equation}
  \begin{array}{lrc cc cc}
                &             &&  CPS' &&   CPS''  \\
   LL|_{P=-1} = &  - VA - AV  &&  +1   &&   +1     \\
   LR|_{P=-1} = &    VA - AV  &&  +1   &&   -1     \\
                &    SP + PS  &&  -1   &&   -1     \\
  \end{array}
\ .
\end{equation}
Therefore, $Q_{X,Y}$ (the $SP+PS$ type)
do not mix with the operators (\ref{eq:chiral_rep1})--(\ref{eq:chiral_rep3})
(the $LL$ and the $LR$ type).

Furthermore, the operators $Q_{7,8}\in ({\bf 8},{\bf 8})$ (the $LR$ type)
do not mix with the $LL$ type operators
($Q_{1,2,3}'\in({\bf 27},{\bf 1}), ({\bf 8},{\bf 1})$),
or with $Q_{5,6}\in({\bf 8},{\bf 1})$ (the $LR$ type)
because the gluon exchange diagrams do not change the flavor structure.

In addition, the mixing between the $({\bf 27},{\bf 1})$
and $({\bf 8},{\bf 1})$ representations
is forbidden by the flavor ${\rm SU}(3)_V$ symmetry.
To sum up, the matrix of the renormalization factor
for the gluon exchange diagrams
has the same structure as in the chiral symmetry preserved case.

Next we investigate the possibility of unwanted mixings
though the penguin diagrams.
In the penguin diagrams for $Q_{7,8} \in ({\bf 8},{\bf 8})$,
a cancellation of the quark loop at the weak operator occurs
between the $d$ quark and the $s$ quark contributions.
This can be seen as the follow.
The penguin diagram for the parity odd part
of the operators $Q_7$, except for the contribution from the spectator quarks,
can be written as
\begin{equation}
   C_7 = C_{VA} - C_{AV}
\ ,
\end{equation}
where
\begin{equation}
  C_{\Gamma_1 \Gamma_2} =
    {\Large {\rm T} [ }
       \, s(X) \, (\bar{s} d)(\bar{u}u - \bar{d}d/2 - \bar{s}s/2 )_{\Gamma_1 \Gamma_2}(x)
                \,\, \bar{d}(Y) \,
    {\Large ] }
\ ,
\end{equation}
at the space-time position $x$,
with the external $s$ quark $s(X)$ at $X$
and the $d$ quark $d(Y)$ at $Y$.
Rewriting with the quark propagator $Q_{q}(x,y)$ for the quark $q$,
we obtain,
\begin{eqnarray}
  C_{\Gamma_1 \Gamma_2}
& = &
   + Q_s(X,x) \,\Gamma_1\, Q_d(x,Y) \,
     {\rm Tr}\left[ \bigl(- Q_u(x,x) + Q_d(x,x)/2 + Q_s(x,x)/2 \bigl) \,\Gamma_2\,
             \right]
\cr
&&
   - Q_s(X,x) \,\Gamma_1\, Q_d(x,x) \,\Gamma_2\, Q_d(x,Y) /2
   - Q_s(X,x) \,\Gamma_2\, Q_s(x,x) \,\Gamma_1\, Q_d(x,Y) /2
\ .
\quad
\end{eqnarray}
Using the isospin symmetry $Q_u=Q_d$,
$C_7$ can be written by
\begin{eqnarray}
  C_7
& = & C_{VA} - C_{AV} \cr
& = &
  \,\,\,\,
  \Bigl[\,\,\,\,
        - Q_s(X,x) \,\gamma_\mu \, Q_d(x,Y) \,
          {\rm Tr}\left[ \bigl( Q_d(x,x) - Q_s(x,x) \bigl) \,\gamma_\mu\gamma_5\,
          \right]/2
\cr
&& \quad\quad
   - Q_s(X,x) \,\gamma_\mu         \, Q_d(x,x) \,\gamma_\mu\gamma_5\, Q_d(x,Y) /2
   - Q_s(X,x) \,\gamma_\mu\gamma_5 \, Q_s(x,x) \,\gamma_\mu        \, Q_d(x,Y) /2
  \,\, \Bigr]
\cr
%
%
&&
 - \Big[ \,
      - Q_s(X,x) \,\gamma_\mu\gamma_5 \, Q_d(x,Y) \,
        {\rm Tr}\left[ \bigl( Q_d(x,x) - Q_s(x,x) \bigl) \,\gamma_\mu\,
              \right]/2
\cr
&& \quad\quad
     - Q_s(X,x) \,\gamma_\mu\gamma_5 \, Q_d(x,x) \,\gamma_\mu         \, Q_d(x,Y) /2
     - Q_s(X,x) \,\gamma_\mu         \, Q_s(x,x) \,\gamma_\mu\gamma_5 \, Q_d(x,Y) /2
  \,\, \Bigr]
\cr
&= &
  -  Q_s(X,x) \,\gamma_\mu \, Q_d(x,Y) \,
     {\rm Tr}\left[ Q_{ds}(x,x) \,\gamma_\mu\gamma_5\, \right]/2
\cr
&&
  +  Q_s(X,x) \,\gamma_\mu\gamma_5 \, Q_d(x,Y) \,
     {\rm Tr}\left[ \bigl( Q_{ds}(x,x) \,\gamma_\mu\, \right]/2
\cr
&&
   - Q_s(X,x) \,\gamma_\mu Q_{ds}(x,x) \,\gamma_\mu\gamma_5\, Q_d(x,Y) /2
\cr
&&
   + Q_s(X,x) \,\gamma_\mu\gamma_5 \, Q_{ds}(x,x) \,\gamma_\mu  \, Q_d(x,Y) /2
\ ,
\end{eqnarray}
where $Q_{ds}(x,x) = Q_d(x,x) - Q_s(x,x)$.
Here we see that
a cancellation of the quark loops at the weak operator occurs
between the $d$ quark and the $s$ quark contributions.
We can see this cancellation also in the operator $Q_8$.

Since this cancellation means that
the renormalization factor coming from the penguin diagram
is proportional to the quark mass difference $(m_d-m_s)$,
mixings to four-fermion operators are absent due to the dimensional reason.
In addition,
the operator arising from the penguin diagrams should have
the flavor structure
$(\bar{s}d)(\bar{u}u + \bar{d}d + \bar{s}s )$,
which is different from that of $Q_{7,8}$.
Thus, operator mixings from $Q_{7,8} \in ({\bf 8},{\bf 8})$
to the other representations and their reverse are absent.
These statements also hold for $Q_{X,Y}$ in (\ref{eq:op_2a})--(\ref{eq:op_2b})
for the same reason,
and the operators $Q_{X,Y}$ are fully isolated in the theory.
Further mixing between the $({\bf 27},{\bf 1})$
and $({\bf 8},{\bf 1})$ representations in the penguin diagrams
is forbidden by the flavor ${\rm SU}(3)_V$ symmetry.
This concludes the proof on the absence of unwanted mixings
among the parity-odd part of dimension 6 operators.

Up to now, we have shown that
the matrix of the renormalization factor
for the parity odd part of the four-fermion operators in
(\ref{eq:chiral_rep1})--(\ref{eq:chiral_rep3})
have the same structure as that in the chiral symmetry preserved case.
Here we consider the mixing to lower dimensional operators.
From $CPS$ symmetry and the equation of motion of the quark,
there is only one operator with the dimension less than 6,
which is
\begin{equation}
   Q_P = ( m_d - m_s ) \cdot \bar{s} \gamma_5 d
\ .
\label{eq:Q_P}
\end{equation}
This operator also appears in the continuum,
but does not yield a nonvanishing contribution to the physical decay amplitudes,
since it is a total derivative operator.
However, this is not valid for the Wilson fermion action
due to chiral symmetry breaking by the Wilson term,
and the operator (\ref{eq:Q_P}) does give a nonzero unphysical contribution
to the amplitudes on the lattice.
This contribution should be subtracted nonperturbatively,
because the mixing coefficient includes
a power divergence due to the lattice cutoff growing as $1/a^2$.
In the present work,
we subtract it by imposing the following condition~\cite{sub:Dawson,sub:Testa},
\begin{equation}
    \langle 0 | \, \overline{Q}_i             \,| K \rangle
 =  \langle 0 | \, Q_i -  \beta_i \cdot Q_P   \,| K \rangle = 0
\ ,
\label{eq:sub_Q}
\end{equation}
for each operator $Q_i$
in (\ref{eq:weak_op1})--(\ref{eq:weak_op5}).
The matrix of the renormalization factor
of the subtracted operators $\overline{Q}_i$
has the same structure as in the chiral symmetry preserved case.

Here, we mention an ambiguity in the subtraction procedure.
Instead of strictly demanding the subtraction condition (\ref{eq:sub_Q}),
we can choose a different subtracted operator,
$\overline{Q}'_i = \overline{Q}_i + \beta'_i \cdot Q_P$,
where $\beta'_i$ is a finite constant depending on the quark masses.
The constants
do not include the power divergence, and they vanish in the chiral limit.
In general, such a finite ambiguity seems
to remain in the final results of the decay amplitude for finite quark masses,
as pointed out in Ref.~\cite{sub:Testa}.
Our case, however, is not a such case for the following reason.
The operator $Q_P$ can be written as
$Q_P= (m_d-m_s)/ (m_d + m_s) \cdot ( \partial_\mu A_\mu - a \overline{X}_A )$
from
the relation of the partially conserved axial vector current
for the Wilson fermion action, 
where $A_\mu$ is the renormalized axial vector current and
$\overline{X}_A$ is the dimension 5 operator
whose matrix element vanishes in the continuum limit.
Thus, a  $\beta'_i \cdot Q_P$ term yields a contribution of form
$\Delta p \cdot C - \langle \pi\pi | a \overline{X}_A | K\rangle \cdot D$
to the decay amplitude with finite constants $C$ and $D$,
where $\Delta p$ is the momentum difference between
the initial and the final state.
These contributions do not include any power divergent parts.
Thus, by taking $\Delta p\to 0$ and the continuum limit,
we can safely estimate the physical value of the decay amplitudes
without suffering from the ambiguity.
%
%
\section{ Method }
\label{Sec: Method}
%
\subsection{ Simulation parameters }
Our calculations are carried out on a subset of
gauge configurations previously generated by PACS-CS Collaboration
with the Iwasaki gauge action and nonperturbatively
$O(a)$-improved Wilson fermion action at $\beta=1.9$
on a $32^3\times 64$ lattice~\cite{conf:PACS-CS}.
The subset corresponds to the hopping parameters
$\kappa_{ud}=0.13770$ for the up and the down quark and
$\kappa_{s }=0.13640$ for the strange quark.
In order to improve the statistics
we further generate gauge configurations
by two runs of the simulation.
The first run uses the same algorithm as employed at the same parameters
in Ref.~\cite{conf:PACS-CS}.
The trajectory length is $\tau=1/4$ and
the dead/alive link method with random parallel translation is used.
The length of MD time, {\it i.e.}, the number of trajectories
multiplied by the trajectory length $\tau$, of this run
is $6,000$ units
as compared to $2,000$ for the original run of Ref.~\cite{conf:PACS-CS}.
The second run does not use the dead/alive link method.
All links are active,
the trajectory length equals $\tau=1$,
and the length of run is also $6,000$ MD time units.
We measure hadron Green's functions
and the decay amplitudes at every $25$ MD time units for both runs.
The total length of the run is $12,000$ MD time units
and the total number of gauge configurations employed
for the measurement is $480$.

We estimate statistical errors by the jackknife method
with bins of $10$ configurations ($250$ MD time units).
The parameters determined from the spectrum analysis
are $a=0.091\ {\rm fm}$ for lattice spacing,
$La=2.91\ {\rm fm}$ for spatial lattice size,
and $m_\pi=275.7(1.5)\,{\rm MeV}$ and $m_K=579.7(1.3)\,{\rm MeV}$
for the pion and the $K$ meson masses.

In the present work, 
we consider the decay in the unphysical kinematics,
where the $K$ meson decay to two zero momentum pions.
The energy difference between the initial $K$ meson
and the final two-pion state is
$\Delta E \equiv m_K-E_{\pi\pi}^I = 21(3)\,{\rm MeV}$ for $I=2$
and $36(18)\,{\rm MeV}$ for $I=0$ on our configurations
as shown in the following section.
In the present work,
we assume that these mismatches of the energy
give only small effects to the decay amplitudes.
%
\subsection{ Time correlation function for $K\to\pi\pi$ }
We extract the matrix element
$\langle K | \overline{Q}_i | \pi\pi ; I \rangle$
from the time correlation function for the $K\to\pi\pi$ process,
\begin{equation}
  G^{I}_{i}(t) = \frac{1}{T} \sum_{\delta=0}^{T-1}
       \langle 0 | \, W_{K^0}(t_K+\delta) \,\, \overline{Q}_{i}(t+\delta) \,\,
                      W_{\pi\pi}^{I}(t_\pi+\delta, t_\pi+1+\delta) \, | 0 \rangle
\ .
\label{eq:G_KPIPI}
\end{equation}
Let us describe various features of this definition one by one.
Firstly, $\overline{Q}_i(t)$ is the subtracted weak operator
at the time slice $t$ defined by
\begin{equation}
  \overline{Q}_i(t) = \sum_{\bf x} \overline{Q}_i({\bf x},t)
\ ,
\end{equation}
with the subtracted operator $\overline{Q}_i({\bf x},t)$
at the space-time position $({\bf x},t)$
defined in (\ref{eq:sub_Q}).

Secondly, the operator $W_{K^0}(t)$ is the wall source for the $K^0$ meson
at the time slice $t$,
\begin{equation}
   W_{K^0}(t) = - \overline{W}_{d}(t) \gamma_5 W_{s}(t)
\ ,
\label{eq:wsource_K}
\\
\end{equation}
with the wall source for the quark $q=u,d,s$,
\begin{eqnarray}
&&           {W}_{q}(t) = \sum_{\bf x}      q ({\bf x}, t) \ , \\
&&  \overline{W}_{q}(t) = \sum_{\bf x} \bar{q}({\bf x}, t) \ .
\end{eqnarray}
We adopt $K^0 = - \bar{d} \gamma_5 s$ as the neutral $K$ meson operator,
so our correlation function has an extra minus from the usual convention.

Thirdly, the operator $W_{\pi\pi}^{I}(t_1,t_2)$ in (\ref{eq:G_KPIPI})
is the wall source
for the two-pion state with the isospin $I$,
which is defined by
\begin{eqnarray}
&&  W_{\pi\pi}^{I=2}(t_1,t_2) =
       \left[ \Bigl(   W_{\pi^0}(t_1) W_{\pi^0}(t_2)
                     + W_{\pi^+}(t_1) W_{\pi^-}(t_2)
              \Bigr)/\sqrt{3}
          + \left( t_1 \leftrightarrow t_2 \right)
       \right]/2
\ ,
\label{eq:W_pipi2}
\\
&&  W_{\pi\pi}^{I=0}(t_1,t_2) =
       \left[ \Bigl(            - W_{\pi^{0}}(t_1) W_{\pi^{0}}(t_2) /\sqrt{2}
                       + \sqrt{2} W_{\pi^{+}}(t_1) W_{\pi^{-}}(t_2)
              \Bigr)/\sqrt{3}
          + \left( t_1 \leftrightarrow t_2 \right)
       \right]/2
\ , \qquad
\label{eq:W_pipi0}
\end{eqnarray}
where $W_{\pi^i}(t)$ is the wall source for $\pi^i$ meson at the time slice $t$,
\begin{eqnarray}
&&    W_{\pi^{+}}(t) = - \overline{W}_d(t) \gamma_5 W_u(t)
\ ,
\label{eq:W_pip}
\\
&&    W_{\pi^{0}}(t) =  \left(\, \overline{W}_u(t) \gamma_5 W_u(t)
                               - \overline{W}_d(t) W_d(t) \, \right)/\sqrt{2}
\ ,
\label{eq:W_pi0}
\\
&&    W_{\pi^{-}}(t) =   \overline{W}_u(t) \gamma_5 W_d(t)
\ .
\label{eq:W_pim}
\end{eqnarray}
The wall source of each pion is separated by one lattice unit according to
$t_1=t_\pi$ and $t_2=t_\pi+1$ in (\ref{eq:G_KPIPI})
to avoid contamination from Fierz-rearranged terms.

We impose the periodic boundary condition in all directions.
The summation over $\delta$,
where $T=64$ denotes the temporal size of the lattice,
is taken in (\ref{eq:G_KPIPI}) to improve the statistics.
The time slice of the $K$ meson is set at $t_K=24$ and
that of the two-pion at $t_\pi=0$.
The gauge configurations are fixed to the Coulomb gauge
at the time slice of the wall source $t_K+\delta$, $t_1+\delta$ and $t_2+\delta$
for each $\delta$.

In the calculation of the the mixing coefficient of the lower dimensional operator,
we rewrite the subtraction of the lower dimensional operator (\ref{eq:sub_Q}) as
\begin{equation}
   \overline{Q}_i = Q_i - \beta_i \cdot Q_P = Q_i - \alpha_i \cdot P
\ ,
\label{eq:sub2_Q}
\end{equation}
by (\ref{eq:Q_P}), with
$P=\bar{s}\gamma_5 d$ and
$\alpha_i = (m_d-m_s) \cdot \beta_i$.
The mixing coefficient $\alpha_i$ in (\ref{eq:sub2_Q})
is obtained from the following ratio of the time correlation function,
\begin{equation}
\alpha_{i} =
  \sum_{\delta_1=0}^{T-1} \langle 0 | \, W_{K^0}(t_K+\delta_1) \, Q_{i}(t+\delta_1) \, | 0 \rangle
  \Bigl/ \,
  \sum_{\delta_2=0}^{T-1} \langle 0 | \, W_{K^0}(t_K+\delta_2) \,     P(t+\delta_2) \, | 0 \rangle
\ ,
\label{eq:calc_alpha}
\end{equation}
in the large $t_K-t$ region,
where $P(t)=\sum_{\bf x} P({\bf x},t)$.
%
\subsection{ Quark contractions for $K\to\pi\pi$ and $K\to 0$ }
\label{Sec: Quark contructions}
\begin{figure}[h]
\vspace{-10mm}
\includegraphics[width=13.5cm]{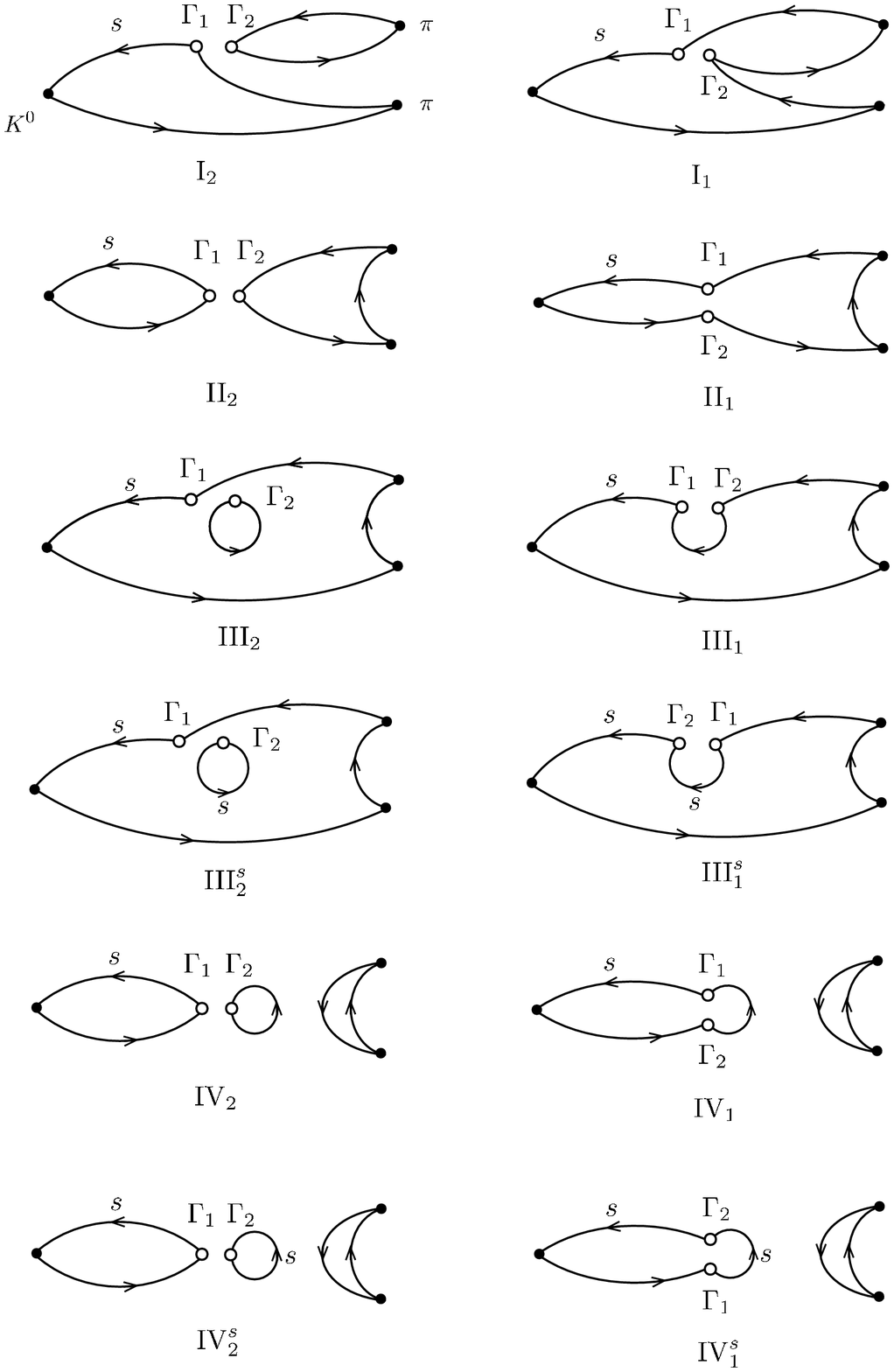}
\caption{
Quark contractions for the time correlation function
for the $K\to\pi\pi$ process
for the operator $Q_i$ ($i=1,2,\cdots 10$).
}
\label{fig:fig_cont_KPIPI}
%
\newpage
\end{figure}

In Fig.~\ref{fig:fig_cont_KPIPI}
we list all of the possible quark contractions
for the $K\to\pi\pi$ time correlation function $G_i^I(t)$ in (\ref{eq:G_KPIPI}).
Again there are a number of features, so let us describe them one by one.
%
\begin{enumerate}
\item
Time runs from right to left in the diagrams.
\item
There are four types of  contractions labeled
${\rm I}$, ${\rm II}$, ${\rm III}$ and ${\rm IV}$.
\item
The diagrams show the quark contractions for the four-fermion operator
\begin{equation}
 Q = \sum_{a,b,c,d}
        (\bar{\psi}^1_a \, \Gamma_1\, \psi^2_b )
        (\bar{\psi}^3_c \, \Gamma_2\, \psi^4_d)
        \, T_{abcd}
\ ,
\end{equation}
with the color indices $a,b,c,d$,
where the spin matrix $\Gamma_{1,2}$
and the color matrix $T_{abcd}$ are given,
depending on the operator $Q_i$, as
\begin{eqnarray}
&&
   \Gamma_1 = \gamma_\mu (1-\gamma_5) \ , \,\,
   \Gamma_2 = \gamma_\mu (1-\gamma_5) \qquad
\mbox{ for $Q_{1,2,3,4,9,10}$ }
\ ,
\label{eq:Gamma12_LL}
\\
&&
   \Gamma_1 = \gamma_\mu (1-\gamma_5) \ , \,\,
   \Gamma_2 = \gamma_\mu (1+\gamma_5) \qquad
\mbox{ for $Q_{5,6,7,8}$ }
\ ,
\label{eq:Gamma12_LR}
\\
&&   T_{abcd} = \delta_{ab} \delta_{cd} \qquad \mbox{ for $Q_{1,3,5,7,9 }$ }
\ ,
\label{eq:T2}  \\
&&   T_{abcd} = \delta_{ad} \delta_{cb} \qquad \mbox{ for $Q_{2,4,6,8,10}$ }
\ .
\label{eq:T1}
\end{eqnarray}
%
\item
In the diagrams,
unmarked line segments represent quark propagators for the $u$ or the $d$ quark,
while those marked by ``$s$'' are for the strange quark.
The filed circles stand for the wall sources for the $K$ meson or pions.
The open circles refer to the matrices $\Gamma_1$ or $\Gamma_2$.
The trace for the spin is taken along closed quark lines.
\item
The subscript $1$ and $2$ attached to the four contraction types
${\rm I}$ though ${\rm IV}$
refers to the number of the trace for the spin.
\item
The superscript ``$s$''
for the contractions ${\rm III}_{1,2}$ and ${\rm IV}_{1,2}$
means that the quark loop at the weak operator
is for the strange quark.
\item
It should be noted that the location of $\Gamma_1$ and $\Gamma_2$
for the contraction ${\rm III}_{1}^{s}$ and ${\rm IV}_{1}^{s}$
are switched from those for ${\rm III}_{1}$ and ${\rm IV}_{1}$.
\item
For the contraction ${\rm IV}_{i}$ and ${\rm IV}^{s}_{i}$ with $i=1,2$
the contribution of the vacuum diagram,
$\langle 0 | K(t_K) Q_i(t) |0\rangle
 \langle 0 | W_{\pi\pi}^{I}(t_\pi,t_\pi+1) | \rangle 0
$,
should be subtracted.
\end{enumerate}

Let us write down some explicit examples.
For the contraction ${\rm I}_2$, we have
\begin{eqnarray}
&&  {\rm I}_2 = \Bigl[\,\,
    \sum_{a,b,c,d}\sum_{\bf x}
      {\rm Tr}\bigl(\, W_d({\bf x},t;t_2) \gamma_5
                       W_d(t_2;{\bf x},t)
                                          \Gamma_2
          \, \bigr)_{dc} \cr
&&
\qquad \times
     {\rm Tr}\bigl(\, W_d({\bf x},t  ;t_1) \gamma_5
                      W_d(        t_1;t_K) \gamma_5
                      W_s(t_K; {\bf x},t )
                                           \Gamma_1
         \, \bigr)_{ba} \cdot T_{abcd}
     \, + \, \bigl(  t_1 \leftrightarrow t_2 \bigr) \,\, \Bigr]/2
\ ,
\label{eq:ex_cont_I22}
\end{eqnarray}
with $t_1=t_\pi$ and $t_2=t_\pi+1$,
where the trace is taken for the spin.
The three types of $W_q$ ($q=d,s$) in (\ref{eq:ex_cont_I22})
are the wall source propagators for the quark $q$
defined by
\begin{eqnarray}
&&    W_q( {\bf x}, t ; t_0 ) = \sum_{\bf y} Q_q( {\bf x}, t ; {\bf y }, t_0 )
\ ,
\label{eq:W_qp_SW}
\\
&&    W_q( t_0 ; {\bf x}, t ) = \sum_{\bf y} Q_q( {\bf y }, t_0; {\bf x}, t  )
                              = \gamma_5 W_q( {\bf x}, t ; t_0 )^{\dagger} \gamma_5
\ ,
\label{eq:W_qp_WS}
\\
&&    W_q( t ; t_0 ) = \sum_{\bf x} W_q( {\bf x}, t ; t_0 )
\ ,
\label{eq:W_qp_WW}
\end{eqnarray}
with the quark propagator $Q_q({\bf x},t ; {\bf y},t_0 )$.
Similarly, the contraction ${\rm I}_1$
is given by
\begin{eqnarray}
&&  {\rm I}_1 = \Biggl[\,\,
    \sum_{a,b,c,d}\sum_{\bf x}
     {\rm Tr}\biggl[\,
                  \bigl(\,
                      W_d({\bf x},t;t_2)   \gamma_5
                      W_d(t_2;{\bf x},t)
                                           \Gamma_2
                   \, \bigr)_{bc} \cr
&&
\qquad \qquad \times
                  \bigl(\,
                      W_d({\bf x},t  ;t_1) \gamma_5
                      W_d(        t_1;t_K) \gamma_5
                      W_s(t_K; {\bf x},t )
                                           \Gamma_1
                  \, \bigr)_{da}
         \, \biggr] \cdot T_{abcd}
    \, + \, \bigl(  t_1 \leftrightarrow t_2 \bigr) \,\, \Biggr]/2
\ ,
\label{eq:ex_cont_I11}
\end{eqnarray}
where the trace is taken for the spin.

The contraction ${\rm II}_2$
is given by
\begin{eqnarray}
&&  {\rm II}_2 = \Bigl[\,\,
    \sum_{a,b,c,d}\sum_{\bf x}
      {\rm Tr}\bigl(\, W_d({\bf x},t;t_2) \gamma_5
                       W_d(t_2      ;t_1) \gamma_5
                       W_d(t_1;{\bf x},t)
                                          \Gamma_2
          \, \bigr)_{dc} \cr
&&
\qquad\qquad\qquad \times
     {\rm Tr}\bigl(\, W_d({\bf x},t  ;t_K) \gamma_5
                      W_s(t_K; {\bf x},t )
                                           \Gamma_1
         \, \bigr)_{ba} \cdot T_{abcd}
     \, + \, \bigl(  t_1 \leftrightarrow t_2 \bigr) \,\, \Bigr]/2
\ ,
\label{eq:ex_cont_II22}
\end{eqnarray}
and
the contraction ${\rm II}_1$
is given by
\begin{eqnarray}
&&  {\rm II}_1 = \Bigl[\,\,
    \sum_{a,b,c,d}\sum_{\bf x}
     {\rm Tr}\biggl[\,
              \bigl(\, W_d({\bf x},t;t_2) \gamma_5
                       W_d(t_2      ;t_1) \gamma_5
                       W_d(t_1;{\bf x},t)
                                          \Gamma_2
          \, \bigr)_{bc} \cr
&&
\qquad\qquad\qquad \times
              \bigl(\, W_d({\bf x},t  ;t_K) \gamma_5
                       W_s(t_K; {\bf x},t )
                                            \Gamma_1
             \, \bigr)_{da}
        \, \biggr] \cdot T_{abcd}
     \, + \, \bigl(  t_1 \leftrightarrow t_2 \bigr) \,\, \Bigr]/2
\ .
\label{eq:ex_cont_II11}
\end{eqnarray}

The contraction ${\rm III}_2$
is given by
\begin{eqnarray}
&&  {\rm III}_2 = \Biggl[\,\,
    \sum_{a,b,c,d}\sum_{\bf x}
     {\rm Tr}\bigl(\,
                      W_d({\bf x},t  ;t_2) \gamma_5
                      W_d(        t_2;t_1) \gamma_5
                      W_d(        t_1;t_K) \gamma_5
                      W_s(t_K; {\bf x},t )
                                           \Gamma_1
         \, \bigr)_{ba} \cr
&&
\qquad\qquad\qquad \times
     {\rm Tr}\bigl(\, Q_d({\bf x},t;{\bf x},t)
                                          \Gamma_2
         \, \bigr)_{dc} \cdot T_{abcd}
   \,  + \, \bigl(  t_1 \leftrightarrow t_2 \bigr) \,\, \Biggr]/2
\ ,
\label{eq:ex_cont_III22}
\end{eqnarray}
where the quark loop for the $d$ quark $Q_d({\bf x},t;{\bf x},t)$
is calculated by
the the stochastic method, whose detail is discussed in the next section.
The contraction ${\rm III}_2^s$ is obtained
by changing $Q_d({\bf x},t;{\bf x},t)$ to
the quark loop for the $s$ quark $Q_s({\bf x},t;{\bf x},t)$.

Having constructed various quark contractions,
we can build the $K\to\pi\pi$ time correlation function
$G_i^I(t)$ for the operators $Q_i$
in the isospin channel $I$ in the following way.
For the $I=2$ case,
we have
\begin{eqnarray}
&&
 G^{I=2}_{1} = \frac{\sqrt{3}}{3} \, \bigl( {\rm I}_2 - {\rm I}_1 \bigr)
 = G^{I=2}_{2}
 = \frac{2}{3} \, G^{I=2}_{9 }
 = \frac{2}{3} \, G^{I=2}_{10}
\ ,
\label{eq:cont_KPIPI_I2_Q12}
\\
&&
 G^{I=2}_{7} = \frac{\sqrt{3}}{2} \, \bigl( {\rm I}_2 - {\rm I}_1 \bigr)
\ ,
\label{eq:cont_KPIPI_I2_Q7}
\\
&&
 G^{I=2}_{8} = \frac{\sqrt{3}}{2} \, \bigl( {\rm I}_2 - {\rm I}_1 \bigr)
\ ,
\label{eq:cont_KPIPI_I2_Q8}
\end{eqnarray}
where $\Gamma_{1,2}$ and $T_{abcd}$ in each contractions
should be chosen according to (\ref{eq:Gamma12_LL})--(\ref{eq:T1})
for each operator.
The relation among different operators (\ref{eq:cont_KPIPI_I2_Q12})
follows from the Fierz identity.

The formulae for the $I=0$ channel are given as follows:
\begin{eqnarray}
&& \mbox{ for \, $(\bar{s}d)(\bar{u}u) = Q_{1,2}$ }   \cr
&& \qquad\qquad
 G^{I=0} = \sqrt{\frac{1}{6}} \,
      \bigl( - {\rm I}_2 - 2\cdot{\rm I}_1
             + 3\cdot{\rm II }_2
             + 3\cdot{\rm T  }_2 \bigr)
\ ,
\label{eq:cont_KPIPI_I01}
\\
&& \mbox{ for \, $(\bar{s}d)(\bar{u}u+\bar{d}d+\bar{s}s) = Q_{3,4,5,6}$ }   \cr
&& \qquad\qquad
 G^{I=0} = \sqrt{\frac{3}{2}} \,
      \bigl( - {\rm I}_2
             + 2\cdot{\rm II }_2   -   {\rm II }_1
             + 2\cdot{\rm T  }_2   -   {\rm T  }_1
             +       {\rm T  }_2^s - {\rm T  }_1^s
      \bigr)
\ ,
\label{eq:cont_KPIPI_I02}
\\
&& \mbox{ for \, $(\bar{s}d)(\bar{u}u-\bar{d}d/2-\bar{s}s/2)
             = Q_{7,8,9,10}$ }   \cr
&& \qquad\qquad
 G^{I=0} = \sqrt{\frac{3}{8}} \,
      \bigl( - {\rm I  }_2   - {\rm I  }_1
             + {\rm II }_2   + {\rm II }_1
             + {\rm T  }_2   + {\rm T  }_1
             - {\rm T  }_2^s + {\rm T  }_1^s
      \bigr)
\ ,
\label{eq:cont_KPIPI_I03}
\end{eqnarray}
with
${\rm T}_i   = {\rm III}_i   - {\rm IV}_i  $ and
${\rm T}_i^s = {\rm III}_i^s - {\rm IV}_i^s$ ($i=1,2$),
where $\Gamma_{1,2}$ and $T_{abcd}$ in each contractions
should be chosen according to (\ref{eq:Gamma12_LL})--(\ref{eq:T1})
for each operator.

\begin{figure}[h]
\includegraphics[width=10.0cm]{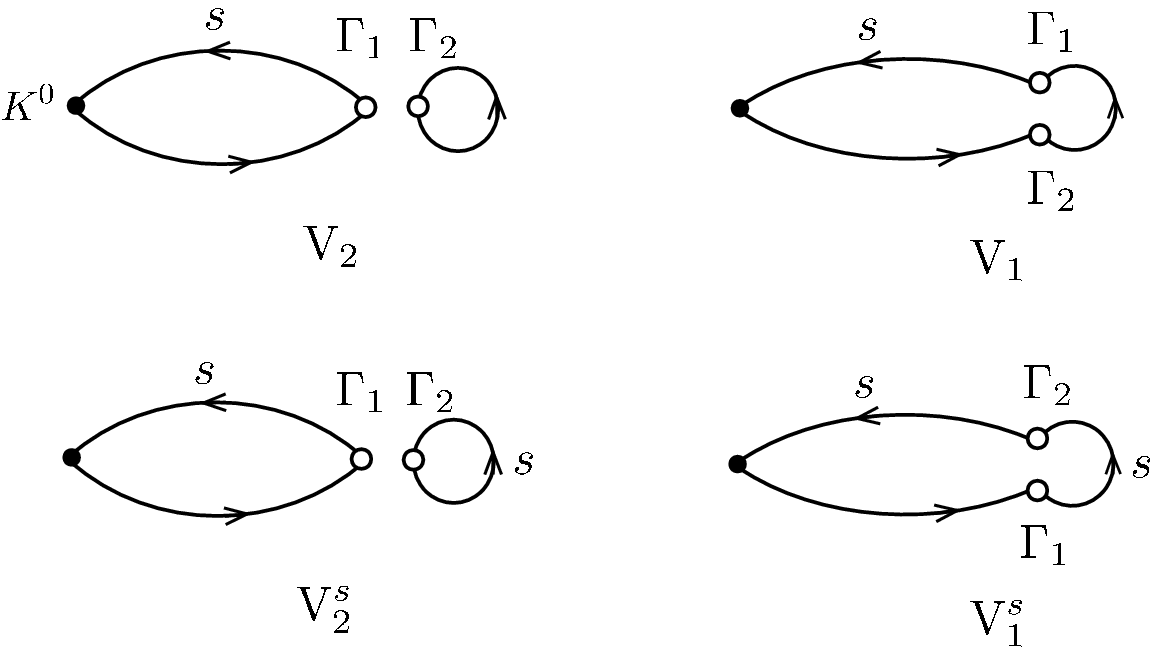}
\caption{
Quark contractions for the time correlation function
for the $K\to 0$ process
for the operator $Q_i$ ($i=1,2,\cdots 10$).
}
\label{fig:fig_cont_K0}
\end{figure}

We now turn to the quark contractions
needed to subtract the contribution
of the lower dimension operator $Q_P$.
In Fig.~\ref{fig:fig_cont_K0}
we list all of the possible quark contractions
for the $K$-to-vacuum time correlation function
$G_{K\to 0} = \langle 0 | W_{K^0}(t_K) Q_i(t) | 0 \rangle$
in (\ref{eq:calc_alpha}).
The notations are the same as for Fig.~\ref{fig:fig_cont_KPIPI}.
The contractions for the operators $Q_i$
are given by
\begin{eqnarray}
&& \mbox{ for \, $(\bar{s}d)(\bar{u}u) = Q_{1,2}$ }   \cr
&& \qquad\qquad
 G_{K\to 0} = - {\rm V}_2
\ ,
\label{eq:cont_K01}
\\
&& \mbox{ for \, $(\bar{s}d)(\bar{u}u+\bar{d}d+\bar{s}s) = Q_{3,4,5,6}$ }   \cr
&& \qquad\qquad
 G_{K\to 0} =  - 2\cdot {\rm V}_2   + {\rm V}_1
      -        {\rm V}_2^s + {\rm V}_1^s
\ ,
\label{eq:cont_K02}
\\
&& \mbox{ for \, $(\bar{s}d)(\bar{u}u-\bar{d}d/2-\bar{s}s/2)
             = Q_{7,8,9,10}$ }   \cr
&& \qquad\qquad
 G_{K\to 0} =  ( - {\rm V}_2   - {\rm V}_1
        + {\rm V}_2^s - {\rm V}_1^s )/2
\ ,
\label{eq:cont_K03}
\end{eqnarray}
where $\Gamma_{1,2}$ and $T_{abcd}$ in each contractions
should be chosen according to (\ref{eq:Gamma12_LL})--(\ref{eq:T1})
for each operator.
We can obtain
the mixing coefficient of the lower dimensional operator
$\alpha_i$ in (\ref{eq:sub2_Q}) by
dividing these by the time correlation function
$\langle 0 | W_{K^0}(t_K) P(t) | 0 \rangle$
as (\ref{eq:calc_alpha}).

\begin{figure}[h]
\includegraphics[width=13.5cm]{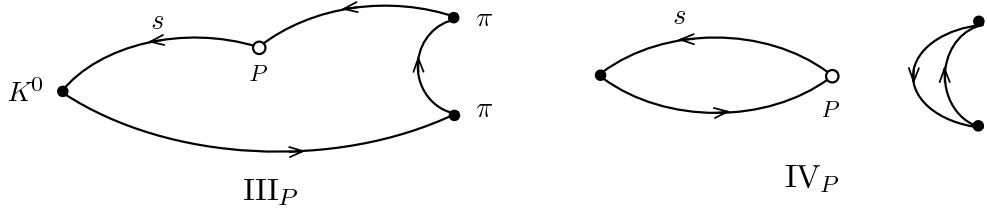}
\caption{
Quark contractions for the time correlation function
for the $K\to\pi\pi$ process
for the operator $P=\bar{s}\gamma_5 d$.
}
\label{fig:fig_cont_KPIPI_P}
\end{figure}

The $K\to \pi\pi$ time correlation function
for the operator $P=\bar{s}\gamma_5 d$
is calculated by
\begin{equation}
 G_P^{I}(t) =
   \frac{1}{T} \sum_{\delta=0}^{T-1}
         \langle 0 | \, W_{K^0}(t_K+\delta) \,\, P(t+\delta) \,\,
                     W_{\pi\pi}^{I}(t_\pi+\delta, t_\pi+1+\delta) \, | 0 \rangle
    = - \frac{3}{\sqrt{6}} \, {\rm T}_P
\ ,
\label{eq:cont_KPIPI_P}
\end{equation}
where ${\rm T}_P = {\rm III}_P - {\rm IV}_P$,
and the contractions ${\rm III}_P$ and ${\rm IV}_P$
are shown in Fig.~\ref{fig:fig_cont_KPIPI_P}.
For the contractions ${\rm IV}_{P}$, 
the contribution of the vacuum diagram
$\langle 0 | K(t_K) P(t) |0\rangle
 \langle 0 | W_{\pi\pi}^{I}(t_\pi,t_\pi+1) | 0 \rangle
$
is supposed to be subtracted.

The $K\to\pi\pi$ time correlation
for the subtracted operator $\overline{Q}_i = Q_i - \alpha_i \cdot P$
is given by subtracting the contributions of $\alpha_i \cdot P$
from those of the operator $Q_i$.
We write these subtractions as
\begin{eqnarray}
&& \rm III \, \to \,
       {\rm III} - \alpha_i \, \frac{-3}{\sqrt{6}} \, {\rm III}_P  \ ,
\label{eq:lowdim_sub_3}
\\
&& \rm IV  \, \to \,
       {\rm IV } - \alpha_i \, \frac{-3}{\sqrt{6}} \, {\rm IV }_P  \ ,
\label{eq:lowdim_sub_4}
\end{eqnarray}
dividing into the connected (${\rm III}$ and ${\rm III}_P$)
and the disconnected contractions (${\rm IV}$ and ${\rm IV}_P $),
where ${\rm III}$ means
the total contribution from the two contractions
${\rm III}_i$ and ${\rm III}^s_i$ for $i=1,2$,
and, similarly, for ${\rm IV}$.
%
\subsection{ Calculation of quark loop }
The quark loop at the weak operator $Q(x,x)$,
{\it i.e.},
the quark propagator starting from the position of the weak operator
and ending at the same position,
appears in the quark contractions
${\rm III}$, ${\rm IV}$, ${\rm III}^s$ and ${\rm IV}^s$
for the $K\to\pi\pi$ process,
and
${\rm V}$ and ${\rm V}^s$ for the $K\to 0$ process,
as shown in the previous subsection.
We calculate them by the stochastic method with
the hopping parameter expansion technique (HPE)
and the truncated solver method (TSM)
proposed in Ref.~\cite{HPE_RSM:Bali}.

The Wilson fermion action can be written as
\begin{equation}
  S^{W}
    = \bar{\psi}\, W        \, \psi
    = \bar{\psi}\, ( M - D )\, \psi
    = \bar{\psi}\, M ( 1 - \bar{D} \, )\, \psi
\ ,
\label{eq:action_W}
\end{equation}
where $\bar{D}= M^{-1} D$ and
\begin{eqnarray}
&&
   ( M \psi )(x) = \left( 1 - \kappa C_{SW} (\sigma \cdot F(x) )/2 \right) \psi (x)
\ ,   \\
&&
   ( D \psi )(x)
   = \left( \sum_\mu \left( D_\mu^+ + D_\mu^- \right) \psi \right)(x)
\ , \\
&&
   ( D_\mu^{+} \psi)(x) = \kappa \, (1-\gamma_\mu)\, U_\mu(x) \psi(x+\mu)
\ ,
\label{eq:D_mu_p}
\\
&&
   ( D_\mu^{-} \psi)(x) = \kappa \, (1+\gamma_\mu) \, U_\mu^\dagger(x-\mu) \psi(x-\mu)
\ .
\label{eq:D_mu_m}
\end{eqnarray}
The quark field of the Wilson fermion $\psi$ is related to
that in the continuum theory $\psi^c$ by $\psi^c = \sqrt{2\kappa}\cdot \psi$
in the tree order.
From (\ref{eq:action_W}) the quark propagator $Q$ can be written
by a hopping parameter expansion form as
\begin{equation}
  Q = W^{-1} = \sum_{n=0}^{\infty} \bar{D}^n M^{-1}
    = \sum_{n=0}^{k-1} \bar{D}^n M^{-1} + \bar{D}^k W^{-1}
\end{equation}
for any integer value of $k$.
We use this expansion to calculate the quark loop $Q(x,x)$ at the weak operator.
In this case, the terms with odd power of $\bar{D}$
do not contribute,
thus
\begin{equation}
   Q(x,x)=( M^{-1} + \bar{D}^2 M^{-1} + \bar{D}^4 W^{-1} )(x,x)
\ ,
\label{eq:Q_ident4}
\end{equation}
for $k=4$.
We can replace $\bar{D}^2$ term by
\begin{equation}
  \bar{D}_{2L} = \sum_{\mu}
                  \, \left( \bar{D}_\mu^{+} \bar{D}_\mu^{-}  +
                            \bar{D}_\mu^{-} \bar{D}_\mu^{+}
                     \right)
\ ,
\label{eq:D2_L}
\end{equation}
with $\bar{D}_\mu^{\pm}= M^{-1} D_\mu^{\pm}$.
Using these expressions,
we calculate the quark loop by the stochastic method according to
\begin{equation}
  Q({\bf x},t ; {\bf x},t ) =
    \frac{1}{N_R}
    \sum_{i=1}^{N_R}\,
         \xi^{*}_i({\bf x},t) \,\, S_i({\bf x},t)
\ .
\label{eq:HPE_stoch}
\end{equation}
The function $S_i({\bf x},t)$ is defined by
\begin{eqnarray}
&&
  S_i({\bf x},t)
   =  \sum_{\bf y} \,
      \bigl( M^{-1} + \bar{D}_{2L} M^{-1} + \bar{D}^4 W^{-1}
      \bigr)({\bf x}, t; {\bf y}, t )
      \,\,  \xi_i({\bf y},t)
\ ,
\label{eq:HPE_stoch_S}
\end{eqnarray}
where
we introduce an $U(1)$ noise $\xi_i({\bf x},t)$ which satisfies
\begin{equation}
  \delta^3 ( {\bf x} - {\bf y} ) =
  \frac{1}{N_R}
  \sum_{i=1}^{N_R}
       \xi_i^{*} ({\bf x},t) \xi_i ({\bf y},t)
\ ,
\label{eq:rand_xi_prop}
\end{equation}
for $N_R \to \infty$.
The effect of HPE for the quark loop
is to remove the $\bar{D}$ and $\bar{D}^3$ terms
in (\ref{eq:HPE_stoch_S}) explicitly
which make only statistical noise.
We find that
HPE reduces the statistical error of the decay amplitudes
to about $50\%$ compared with the normal stochastic method.

We also implement the truncated solver method (TSM)
for (\ref{eq:HPE_stoch}) by
\begin{equation}
  Q({\bf x},t ; {\bf x},t ) =
  \frac{1}{N_R}
  \sum_{i=1}^{N_R}\,
       \xi^{*}_i({\bf x},t) \,\, \bigl[ S_i({\bf x},t) - S_i^T({\bf x},t) \bigr]
  \,\,
  +
  \,\,
  \frac{1}{N_T}
  \sum_{i=N_R+1}^{N_R+N_T}\,
       \xi^{*}_i({\bf x},t) \,\, S_i^T({\bf x},t)
\ ,
\label{eq:TSM}
\end{equation}
where
$S^T_i({\bf x},t)$ is a value given with the quark propagator $W^{-1}$
in (\ref{eq:HPE_stoch_S}) calculated with a loose stopping condition,
and $S_i({\bf x},t)$ is that with a stringent condition.
We set $N_T=5$ and the stopping condition
$R\equiv | W x  - \xi |/|\xi| < 1.2 \times 10^{-6}$
with $x$ denoting the iterative solution of
$W x = \xi$ for $S_i^T({\bf x},t)$,
and $N_R=1$ and $R < 10^{-14}$ for $S_i({\bf x},t)$ in the present work.
The numerical cost of TSM (\ref{eq:TSM})
is about twice of that without TSM (\ref{eq:HPE_stoch}) with $N_R=1$.
%
\subsection{ Time correlation function for $\pi\pi\to\pi\pi$ }
We calculate two types of time correlation functions for $\pi\pi\to\pi\pi$
to obtain the normalization factors
which are  needed to extract the matrix elements
$\langle K | \overline{Q}_i | \pi\pi ; I \rangle$
from the time correlation function
$G^{I}_{i}(t)$ in (\ref{eq:G_KPIPI}).
These are point-wall and wall-wall time correlation functions,
which are defined by
\begin{eqnarray}
&&  G_{PW}^{I}(t) = \frac{1}{T} \sum_{\delta=0}^{T-1}
             \langle 0 | \, (\pi\pi)^{I}(t+\delta)\,\,
                            W_{\pi\pi}^{I}(t_\pi+\delta, t_\pi+1+\delta) \, | 0 \rangle
\ ,
\label{eq:G_PIPI_PW}
\\
&&  G_{WW}^{I}(t) = \frac{1}{T} \sum_{\delta=0}^{T-1}
             \langle 0 | \, W_{\pi\pi}^{I}(t    +\delta, t    +1+\delta) \,\,
                            W_{\pi\pi}^{I}(t_\pi+\delta, t_\pi+1+\delta) \, | 0 \rangle
\ ,
\label{eq:G_PIPI_WW}
\end{eqnarray}
where $(\pi\pi)^{I}(t)$ is the operator for the two-pion system with the isospin $I$,
\begin{eqnarray}
&&  (\pi\pi)^{I=2}(t) =
      \sum_{{\bf x},{\bf y}}
         \Bigl(   \pi^0({\bf x},t) \pi^0({\bf y},t)
                + \pi^+({\bf x},t) \pi^-({\bf y},t)
         \Bigr)/\sqrt{3}
\ ,
\label{eq:psource_pipi2}
\\
&&  (\pi\pi)^{I=0}(t) =
      \sum_{{\bf x},{\bf y}}
         \Bigl( -          \pi^0({\bf x},t) \pi^0({\bf y},t) /\sqrt{2}
                + \sqrt{2} \pi^+({\bf x},t) \pi^-({\bf y},t)
      \Bigr)/\sqrt{3}
\ ,
\label{eq:psource_pipi0}
\end{eqnarray}
with the operator for $\pi^i$ meson $\pi^i({\bf x},t)$ defined by
\begin{eqnarray}
&&    \pi^{+}({\bf x},t) = -          \bar{d}({\bf x},t) \gamma_5 u({\bf x},t)
\ ,
\label{eq:P_pip}
\\
&&    \pi^{0}({\bf x},t) =  \left(\,  \bar{u}({\bf x},t) \gamma_5 u({\bf x},t)
                                    - \bar{d}({\bf x},t) \gamma_5 d({\bf x},t)
                         \, \right)/\sqrt{2}
\ ,
\label{eq:P_pi0}
\\
&&    \pi^{-}({\bf x},t) =            \bar{u}({\bf x},t) \gamma_5 d({\bf x},t)
\ .
\label{eq:P_pim}
\end{eqnarray}
The operator $W_{\pi\pi}^{I}(t_1,t_2)$ ($I=0,2$) is defined by
(\ref{eq:W_pipi2}) and (\ref{eq:W_pipi0}).
In the present work
we set $t_\pi=0$ in (\ref{eq:G_PIPI_PW}) and (\ref{eq:G_PIPI_WW}).

\begin{figure}[h]
\includegraphics[width=10.0cm]{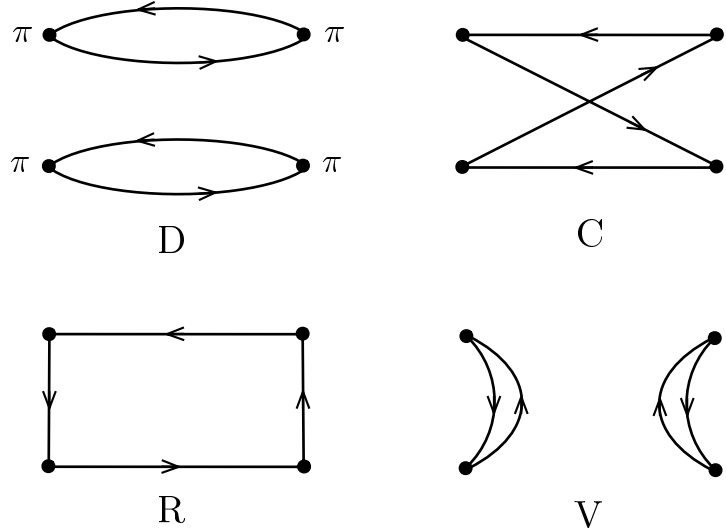}
\caption{
Quark contractions for the time correlation function
for $\pi\pi\to\pi\pi$.
}
\label{fig:fig_cont_PIPI}
\end{figure}

In Fig.~\ref{fig:fig_cont_PIPI},
we list all of the possible quark contractions for the time correlation function
for the $\pi\pi\to\pi\pi$ processes.
Time runs from right to left in the diagrams.
There are four types of contractions,
${\rm D}$, ${\rm C}$, ${\rm R}$ and ${\rm V}$.
The filled circles represent
the wall source $W_{\pi^i}$ in (\ref{eq:W_pip})-(\ref{eq:W_pim})
or the point source $\pi^i$ in (\ref{eq:P_pip})-(\ref{eq:P_pim}) for the pion.
For the contraction ${\rm V}$
the contribution of the vacuum diagram,
$\langle 0 | (\pi\pi)^{I}(t) | 0 \rangle
 \langle 0 | W_{\pi\pi}^{I}(t_\pi,t_\pi+1) | 0 \rangle$
or
$\langle 0 | W_{\pi\pi}^{I}(t    ,t    +1) | 0 \rangle
 \langle 0 | W_{\pi\pi}^{I}(t_\pi,t_\pi+1) | 0 \rangle$,
is supposed to be subtracted.

For example,  the explicit form for the contraction ${\rm C}$
for the point-wall time correlation function
$G_{PW}^{I}(t)$ in (\ref{eq:G_PIPI_PW}) is given by
\begin{eqnarray}
&&
   {\rm C} =  \Bigl[\,\,
          \sum_{{\bf x},{\bf y}}
           {\rm Tr}\bigl(\,
                         W_d(t_1 ; {\bf y},t) \gamma_5
                         W_d({\bf y},t ; t_2) \gamma_5
                         W_d(t_2 ; {\bf x},t) \gamma_5
                         W_d({\bf x},t ; t_1) \gamma_5
                   \, \bigr)
\cr
&&
\qquad\qquad
\qquad\qquad
\qquad\qquad
            \, + \, \bigl(  t_1 \leftrightarrow t_2 \bigr)
         \,\, \Bigr]/2
\ ,
\label{eq:ex_cont_PIPI_PW_C}
\end{eqnarray}
with $t_1=t_\pi$ and $t_2=t_\pi+1$,
and that for the wall-wall correlation function $G_{WW}^{I}(t)$
in (\ref{eq:G_PIPI_WW}) by
\begin{eqnarray}
&&
 {\rm C} =  \Bigl[\,\,
           {\rm Tr}\bigl(\, W_d(t_1 ; t_4 ) \gamma_5
                            W_d(t_4 ; t_2 ) \gamma_5
                            W_d(t_2 ; t_3 ) \gamma_5
                            W_d(t_3 ; t_1 ) \gamma_5
                   \, \bigr)
\cr
&&
\qquad\qquad
\qquad\qquad
\qquad\qquad
            \, + \, \bigl(  t_1 \leftrightarrow t_2 \bigr)
            \, + \, \bigl(  t_3 \leftrightarrow t_4 \bigr)
            \, + \, \bigl(  t_1 \leftrightarrow t_2 ,
                            t_3 \leftrightarrow t_4 \bigr)
         \,\, \Bigr]/4
\ ,
\label{eq:ex_cont_PIPI_WW_C}
\end{eqnarray}
with
$t_1=t_\pi$, $t_2=t_\pi+1$,
$t_3=t    $, $t_4=t    +1$,
where the trace is taken for the color and the spin indices.
The wall source propagators $W_d$ are defined by
(\ref{eq:W_qp_SW})-(\ref{eq:W_qp_WW}).

For the calculation of the contraction ${\rm R}$
for the point-wall time correlation function
$G_{PW}^{I}(t)$ in (\ref{eq:G_PIPI_PW}),
we use the stochastic method according to
\begin{eqnarray}
&&
  {\rm R} =  \Bigl[\,\,
          \frac{1}{N_R} \sum_{i=1}^{N_R}
          \sum_{{\bf x},{\bf y}}
           {\rm Tr}\bigl(\,
                         W_d(t_1 ; t_2      )   \gamma_5
                         W_d(t_2 ; {\bf x},t)   \gamma_5
                      \, Z_i({\bf x},t) \xi_i^{*}({\bf y},t) \, \gamma_5
                         W_d({\bf y},t ; t_1) \gamma_5
                   \, \bigr)
\cr
&&
\qquad\qquad
\qquad\qquad
\qquad\qquad
            \, + \, \bigl(  t_1 \leftrightarrow t_2 \bigr)
         \,\, \Bigr]/2
\ ,
\label{eq:ex_cont_PIPI_PW_R}
\end{eqnarray}
with $t_1=t_\pi$ and $t_2=t_\pi+1$,
where $\xi_i({\bf y},t)$ is an $U(1)$ noise
which satisfies (\ref{eq:rand_xi_prop}),
and $Z_i({\bf x},t)$ is defined by
\begin{equation}
  Z_i({\bf x},t)
   =  \sum_{\bf y} \,
          W^{-1}({\bf x}, t; {\bf y}, t )
      \,\,  \xi_i({\bf y},t)
\ ,
\end{equation}
with the kernel of the Wilson fermion $W$ in (\ref{eq:action_W}).
The contraction ${\rm V}$
is also calculated by using $Z_i({\bf x},t)$.
In actual calculations,
we find that
relaxing the stopping condition to
$| W x  - \xi |/|\xi| < 1.2 \times 10^{-6}$
for the calculation of $Z_i({\bf x},t)$
makes only negligible effects
to the final result, compared with the statistical error.
Thus, we adopt this loose stopping condition
with $N_R=6$ in (\ref{eq:ex_cont_PIPI_PW_R}).

The quark contraction
for the time correlation function,
$G_{PW}^{I}(t)$ or $G_{WW}^{I}(t)$, is given by
\begin{eqnarray}
&&    G^{I=2} = {\rm D} - {\rm C} \ , \\
&&    G^{I=0} = {\rm D} + \frac{1}{2}\, {\rm C} - 3\, {\rm R} + \frac{3}{2}\, {\rm V} \ .
\end{eqnarray}
%
%
\section{ Results }
\label{Sec: Results}
%
\subsection{ Time correlation function for $\pi\pi\to\pi\pi$ }
\label{Sec: Time correlation function of pipi}
\begin{figure}[h]
\includegraphics[height=6.5cm]{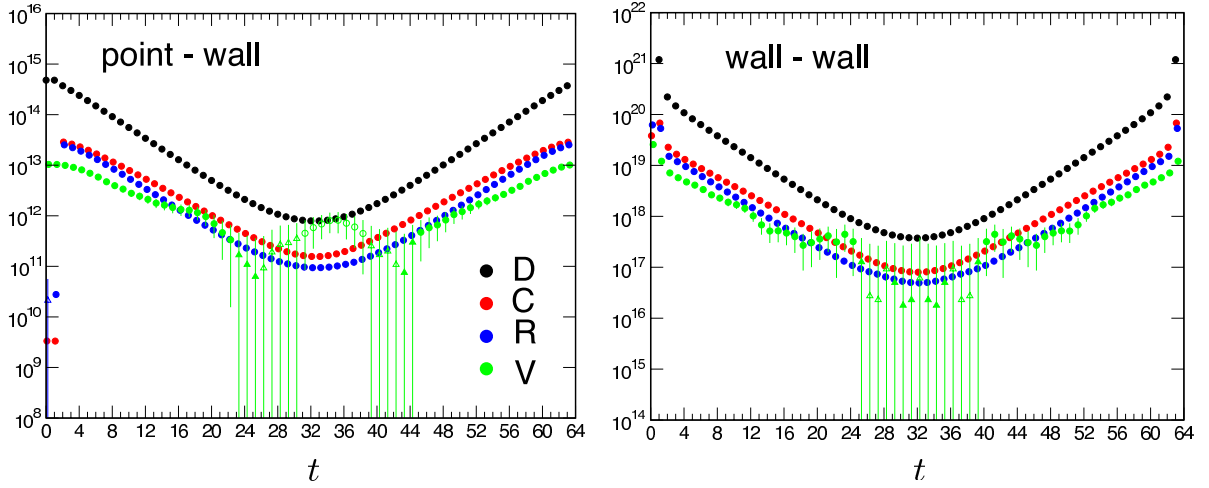}
\caption{
Four types of contractions for
the time correlation function for $\pi\pi\to\pi\pi$.
Left panel shows those for the point-wall function $G_{PW}^{I}(t)$
in (\ref{eq:G_PIPI_PW})
and right for the wall-wall function $G_{WW}^{I}(t)$
in (\ref{eq:G_PIPI_WW}).
}
\label{fig:PIPI_LW_WW}
\end{figure}

Fig.~\ref{fig:PIPI_LW_WW} shows
the contributions of
the four types of contractions, ${\rm D},{\rm C},{\rm R},{\rm V}$,
for the time correlation function for $\pi\pi\to\pi\pi$, with
those for the point-wall function $G_{PW}^{I}(t)$ in (\ref{eq:G_PIPI_PW})
plotted on the left and
those for the wall-wall function $G_{WW}^{I}(t)$ in (\ref{eq:G_PIPI_WW})
on the right panel.
The source operator is placed at $t_\pi=0$.
The time correlation functions
behave in the large time region as
\begin{eqnarray}
&&  G_{PW}^{I}(t) = A^I \cdot
                        \left(   {\rm e}^{ - E_{\pi\pi}^{I} \, t  }
                               + {\rm e}^{ - E_{\pi\pi}^{I} ( T - t )} \right)
                   + C^I
\ ,
\label{eq:fit_G_PIPI_PW}
\\
&&  G_{WW}^{I}(t) = A_{\pi\pi}^{I} \cdot
                         \left(   {\rm e}^{ - E_{\pi\pi}^{I} \, t  }
                                + {\rm e}^{ - E_{\pi\pi}^{I} ( T - t )} \right)
                + D^I
\ ,
\label{eq:fit_G_PIPI_WW}
\end{eqnarray}
where
$E_{\pi\pi}^{I}$ is the energy of the two-pion system with the isospin $I$,
$A^I$ is a constant whose form is irrelevant,
and
\begin{equation}
   A_{\pi\pi}^I
     =    \langle 0 |  W_{\pi\pi}^{I}(0,1) | \pi\pi; I \rangle^2
        / \langle \pi\pi; I | \pi\pi; I \rangle
\ .
\label{eq:A_PIPI}
\end{equation}
The constant terms $C^I$ and $D^I$
in (\ref{eq:fit_G_PIPI_PW}) and (\ref{eq:fit_G_PIPI_WW})
come from the two pions propagating in the opposite time directions
({\it i.e.,} around-the-world effect for the two-pion operator).

\begin{figure}[h]
\includegraphics[height=6.5cm]{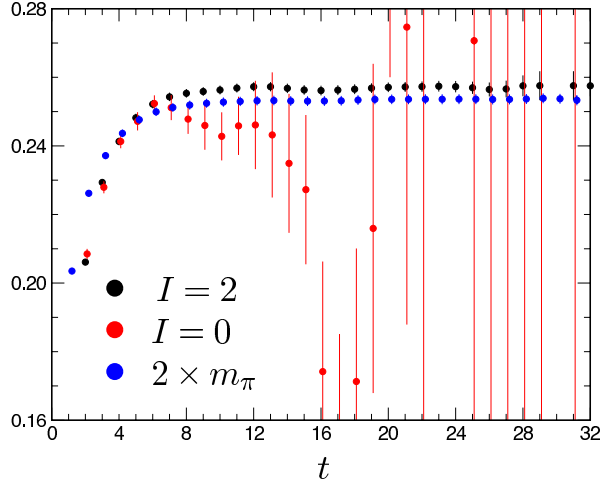}
\caption{
Effective mass of the time correlation function
$G_{PW}^{I}(t)$ for $\pi\pi\to\pi\pi$
with the isospin $I=0$ and $I=2$.
Twice of the effective mass for the pion
is also plotted for a comparison.
}
\label{fig:LM_PIPI02_LW}
%
\newpage
\end{figure}

The effective mass of the point-wall time correlation function $G_{PW}^{I}(t)$
is plotted in Fig.~\ref{fig:LM_PIPI02_LW},
where the effective mass $m_{eff}$ at $t$ is given by
\begin{equation}
     \frac{  G_{PW}^{I}( t + 1 ) - G_{PW}^{I}( t + 4 ) }
          {  G_{PW}^{I}( t     ) - G_{PW}^{I}( t + 3 ) }
  =  \frac{  f( t+1 ; m_{eff} ) - f( t+4 ; m_{eff} ) }
          {  f( t   ; m_{eff} ) - f( t+3 ; m_{eff} ) }
\end{equation}
with the function
$f(t;m_{eff}) = {\rm exp}( - m_{eff} \cdot t ) + {\rm exp}( -m_{eff} \cdot ( T-t) )$.
We find plateaus in the time region $t \ge 9$ for both $I=0$ and $2$,
albeit admittedly much noisier for $I=0$ than for $I=2$.
Compared with the value $2 m_\pi$ plotted in blue,
the two-pion energy for $I=2$ is larger,
signifying repulsive interaction of the two-pion system,
whereas that for $I=0$ is smaller showing attractive interaction.

In the extraction of
the matrix elements
$\langle K | \overline{Q}_i | \pi\pi ; I \rangle$ from the time correlation function
$G^{I}_{i}(t)$ in (\ref{eq:G_KPIPI}),
the values of $E_{\pi\pi}^{I}$ and $A_{\pi\pi}^I$ are needed.
Since the statistical error of the point-wall correlation function $G_{PW}^{I}(t)$ is smaller than
that for the wall-wall function $G_{WW}^{I}(t)$,
we first extract the energy $E_{\pi\pi}^{I}$ from $G_{PW}^{I}(t)$,
and then extract the amplitude $A_{\pi\pi}^I$ from $G_{WW}^{I}(t)$
by fitting to (\ref{eq:fit_G_PIPI_WW})
with the determined value of $E_{\pi\pi}^{I}$
and regarding $A_{\pi\pi}^I$ and $D^I$ as unknown parameters.
The results for $E_{\pi\pi}^{I}$ and $A_{\pi\pi}^{I}$ are
\begin{eqnarray}
&&   E_{\pi\pi}^{I=2} = 0.2567(14)                 \ , \qquad
     A_{\pi\pi}^{I=2} = 2.513(27) \times 10^{20}   \ , \cr
&&   E_{\pi\pi}^{I=0} = 0.2499(83)                 \ , \qquad
     A_{\pi\pi}^{I=0} = 2.41(13)  \times 10^{20}   \ ,
\label{eq:table_PIPI_E}
\end{eqnarray}
in the lattice unit,
where we adopt the fitting range
$t=[9,32]$ for $I=2$ and
$t=[9,12]$ for $I=0$.

The mass of the pion and the $K$ meson
obtained in the present work are
$m_\pi = 0.12671(71)$ and $m_K = 0.26641(58)$ in the lattice unit.
The energy difference between the initial $K$ meson and the final two-pion state,
$\Delta E^I = m_K - E_{\pi\pi}^I$,
is
$\Delta E^{I=2} = 0.0097(14)$ ($21(3) \,{\rm MeV}$) and
$\Delta E^{I=0} = 0.0165(83)$ ($36(18)\,{\rm MeV}$).
In the present work,
we assume that these violations of energy conservation
yield only small effects to the results for the $K\to\pi\pi$ decay amplitudes.
%
\subsection{ Time correlation function for $K\to\pi\pi$ in the $I=0$ channel }
\label{Sec: Time correlation function of KPIPI for Delta I=1/2 process}
\begin{figure}[h]
\includegraphics[height=12.0cm]{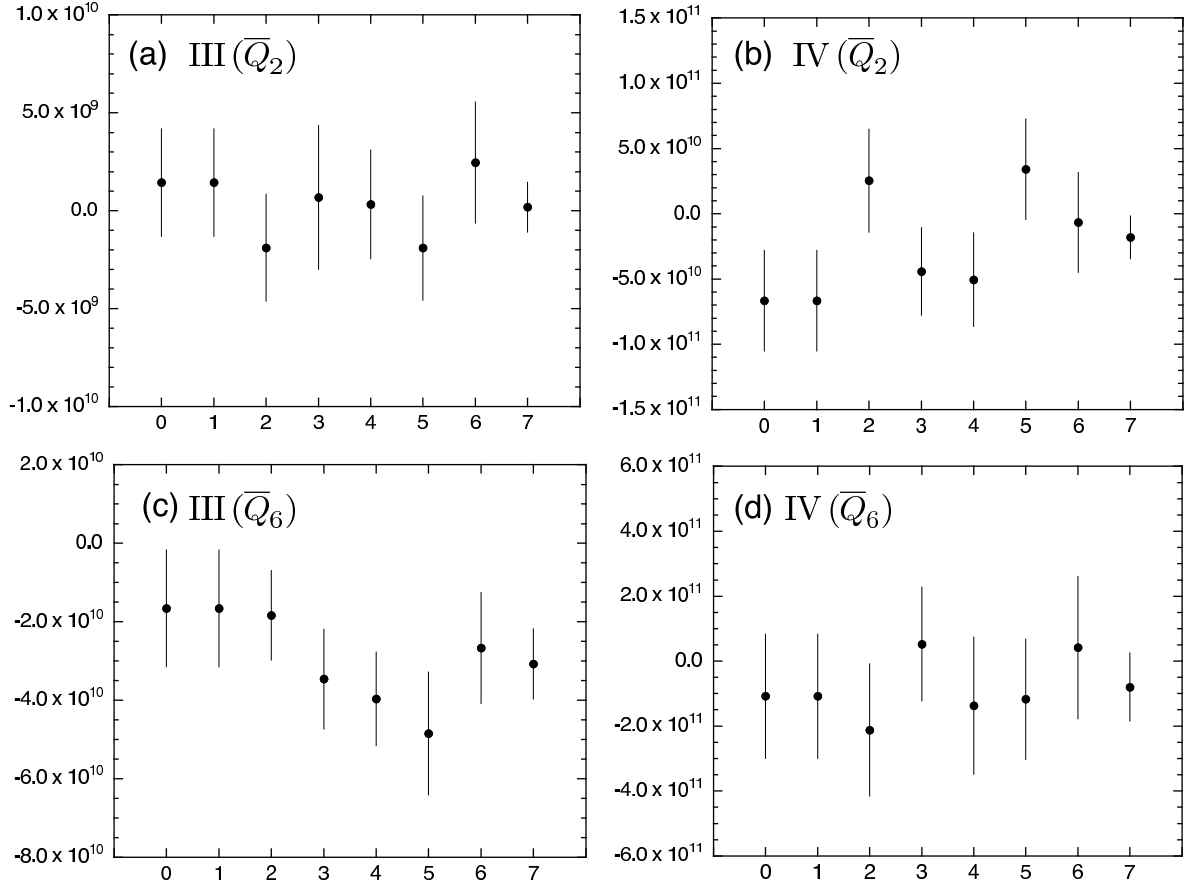}
\caption{
Effect of the truncated solver method.
Panels (a)-(d) show the contributions of the contraction
${\rm III}$ and ${\rm IV}$
to the time correlation functions for $\overline{Q}_2$ and $\overline{Q}_6$ at $t=9$.
In each panel,
the data at $x=0$ show the contribution obtained by
the usual stochastic method (\ref{eq:HPE_stoch}) with $N_R=1$,
{\it i.e.}, the contribution given by setting the quark loop
$Q({\bf x},t ; {\bf x},t ) = \xi^{*}_i({\bf x},t) \, S_i({\bf x},t)$
for $i=1$.
The data at $x=1,2,\cdots , 6$ correspond to
the contributions given by setting
$Q({\bf x},t ; {\bf x},t ) = \xi^{*}_i({\bf x},t) \, S_i^T({\bf x},t)$
for $x=i=1,2,\cdots , 6$ ($N_R+N_T=1+5$) with
$S_i^T({\bf x},t)$ in (\ref{eq:TSM}).
The data at $x=7$ are
average of the data at $x=1,2,\cdots , 6$.
}
\label{fig:TSM_Q2Q6_09.eps}
%
\newpage
\end{figure}

In Fig.~\ref{fig:TSM_Q2Q6_09.eps}
we demonstrate the effects of the truncated solver method.
The four panels (a)--(d) show the contributions of the contractions
${\rm III}$ and ${\rm IV}$
to the time correlation functions for
$\overline{Q}_2$ and $\overline{Q}_6$ at $t=9$.
In each panel,
the data at $x=0$ shows the result of a stochastic estimate
with a stringent stopping condition,
while those at $x=1, \cdots, 6$ are obtained with a loose stopping condition,
with an identical noise vector employed for $x=0$ and $x=1$.
Thus, the difference between the data at $x = 0$ and $x = 1$ corresponds to
the first term of (\ref{eq:TSM}) for $N_R=1$, and
the data at $x=2,\cdots , 6$ ($N_T=5, N_T+N_R=6$ )
correspond to the components of the second term of (\ref{eq:TSM}).
We find that the first term is negligible
compared with the statistical error for all channels.
Thus we can neglect it,
and estimate the quark loop contribution by only the second term as
\begin{equation}
  Q({\bf x},t ; {\bf x},t ) =
  \frac{1}{N_T+N_R}
  \sum_{i=1}^{N_T+N_R}\,
       \xi^{*}_i({\bf x},t) \,\, S_i^T({\bf x},t)
\ .
\label{eq:TSM2}
\end{equation}
The contribution given by the sum (\ref{eq:TSM2})
is plotted at $x=7$ in each panel.
We see that the statistics is significantly improved
by increasing the number of random numbers from $1$ to $6$.

\begin{figure}[h]
\includegraphics[height=12.0cm]{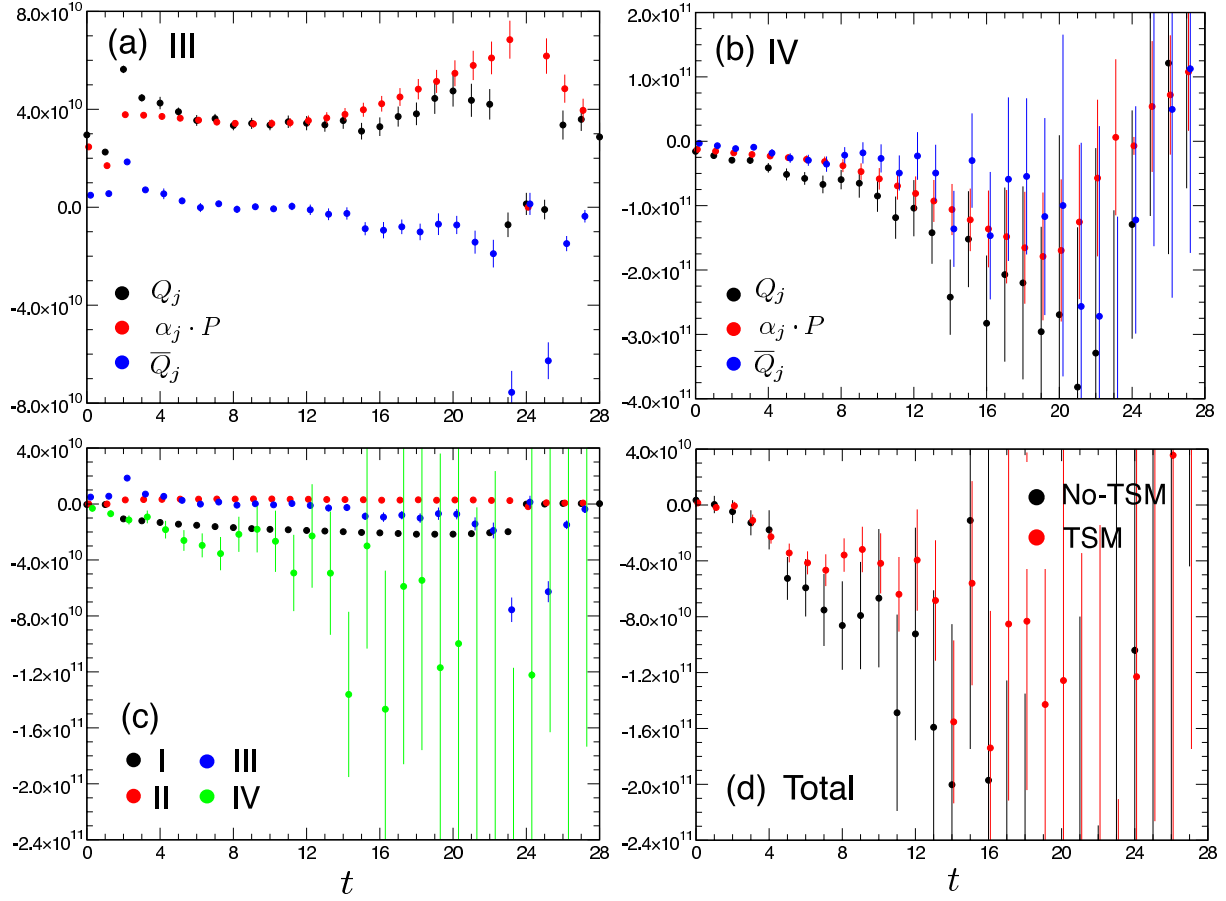}
\caption{
Time correlation function for the operator $Q_2$
for the $\Delta I=1/2$ $K\to\pi\pi$ process,
$G_2^{I=0}(t)$ in (\ref{eq:G_KPIPI}).
The time slices of the two-pion and the $K$ meson
are set at $t_\pi=0$ and $t_K=24$,
while the operator $Q_i$ runs over the whole time extent.
(a) Contributions
    of the contraction ${\rm III}$ for $Q_2$, $\alpha_2\cdot P$ and
    $\overline{Q}_2 = Q_2 - \alpha_2 \cdot P$.
(b) Contributions
    of the contraction ${\rm IV}$ for $Q_2$, $\alpha_2\cdot P$ and
    $\overline{Q}_2 = Q_2 - \alpha_2 \cdot P$.
(c) Contributions from each type of contractions for $\overline{Q}_2$,
(d) Total correlation functions calculated with TSM and without TSM.
}
%
\newpage
\label{fig:G_2_I0}
\end{figure}

The results for the $I=0$ $K\to\pi\pi$ time correlation function
for the operator $Q_2$ ( $G_2^{I=0}(t)$ in (\ref{eq:G_KPIPI}) )
are plotted in Fig.~\ref{fig:G_2_I0}.
The time slice of the two-pion is set at $t_\pi=0$ and the $K$ meson at $t_K=24$,
while the operator $Q_i(t)$ runs over the whole time extent as explained before.
In the panels (a) and (b), we observe a large cancellation
between the contributions from the operator $Q_2$ and the subtraction term $\alpha_2\cdot P$
for both contractions ${\rm III}$ and ${\rm IV}$.
In panel (c) we find that the contribution from the contraction ${\rm IV}$
is similar in magnitude to that from the contraction ${\rm I}$.
This appears different from the previous work
by RBC-UKQCD Collaboration
with the domain wall fermion action
in Refs.~\cite{A0:RBC-UKQCD_400,A0:RBC-UKQCD_300}.
In panel (d), we compare the correlation functions
calculated with TSM and without TSM.
We find that TSM significantly improves the statistics.
The numerical cost of TSM is about twice of that without TSM.
Thus TSM is a very efficient method.

\begin{figure}[h]
\includegraphics[height=12.0cm]{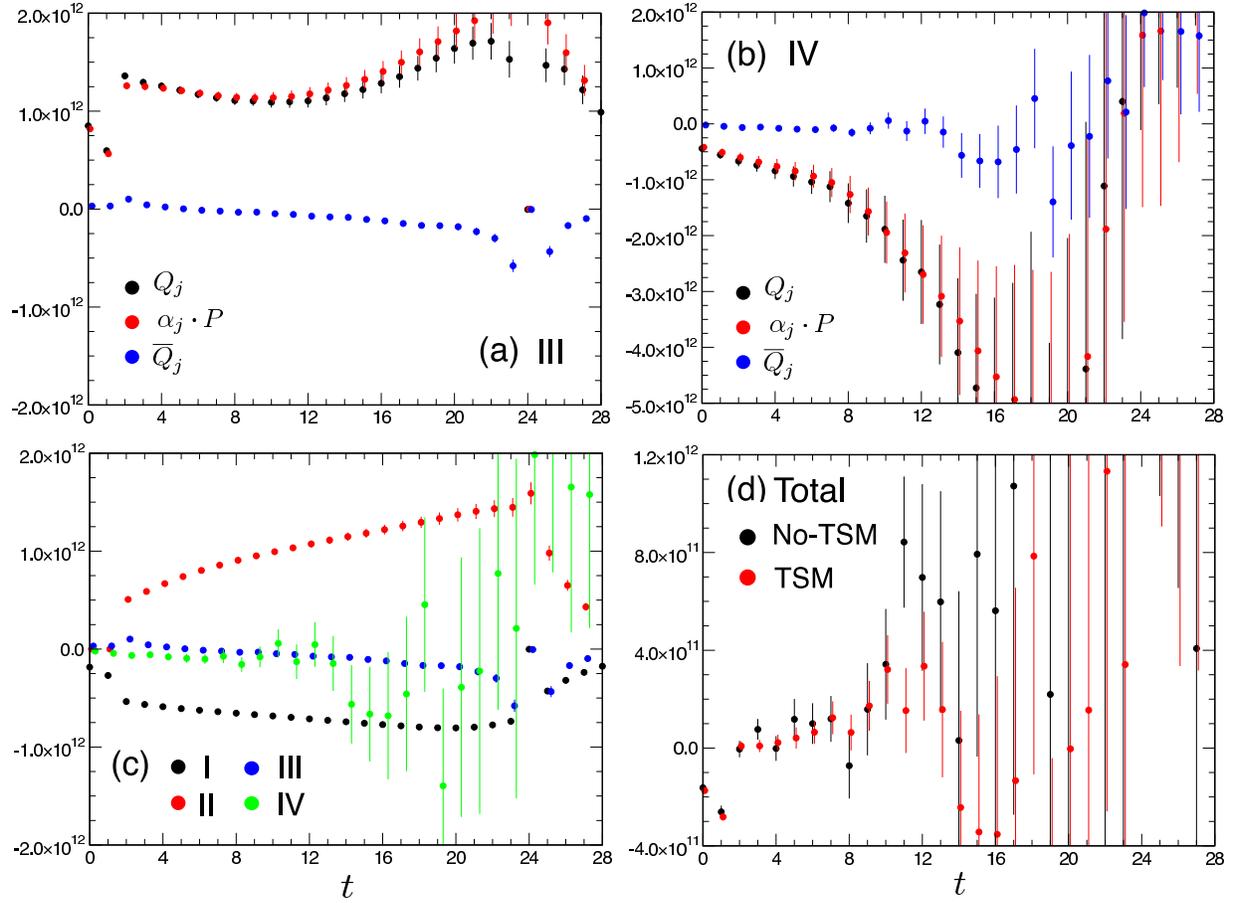}
\caption{
Time correlation function for the operator $Q_6$
for the $\Delta I=1/2$ $K\to\pi\pi$ process,
$G_6^{I=0}(t)$ in (\ref{eq:G_KPIPI}),
following the same convention as in Fig.~\protect\ref{fig:G_2_I0}.
}
\label{fig:G_6_I0}
%
\newpage
\end{figure}

The results for $Q_6$ in the $I=0$ channel are plotted in Fig.~\ref{fig:G_6_I0}.
Here, also, we find a large cancellation between
the contributions of $Q_6$ and the subtraction $\alpha_6\cdot P$
for both contractions ${\rm III}$ and ${\rm IV}$ (see panels (a) and (b)).
In panel (c) a large cancellation is observed  between
the contraction ${\rm I}$ and ${\rm II}$,
which is not the case for $\overline{Q}_2$.
An efficiency of TSM is observed also for $Q_6$ in panel (d).
%
\subsection{ $K\to\pi\pi$ matrix elements }
\label{Sec: Matrix elements}
In order to extract the $K\to\pi\pi$ matrix element,
we consider an effective matrix element $M^{I}_{i}(t)$,
which behaves as
$M^{I}_{i}(t) = M^I_i \equiv \langle K |\, \overline{Q}_i({\bf 0},0)\, | \pi\pi; I \rangle$
in the time region $t_K \gg t \gg t_\pi$.
It can be constructed from the time correlation function
$G^{I}_{i}(t)$ in (\ref{eq:G_KPIPI}) by
\begin{equation}
  M^{I}_{i}(t) = G^{I}_{i}(t)
   / \sqrt{ A_K A_{\pi\pi}^{I} }
   \cdot F^{I}
   \cdot {\rm e}^{ m_K (t_K-t) + E^{I}_{\pi\pi} (t-t_\pi) }
   \times (-1)
\ .
\label{eq:effective_amp}
\end{equation}
Here, the $K$ meson mass $m_K$
and the energy of the two-pion state $E^{I}_{\pi\pi}$ are fixed at the values
obtained from the correlation function of the $K$ meson
and the $\pi\pi\to\pi\pi$.
The factor $(-1)$ comes from the convention of the $K^0$ operator
in (\ref{eq:wsource_K}).
The constant
$A_K = \langle 0 | W_K | K \rangle^2 / \langle K | K \rangle$
is estimated from the wall-wall propagator of the $K$ meson,
with the value $A_K=8.949(34)\times 10^{9}$ in the lattice unit.
The constant $A_{\pi\pi}^{I}$ is defined by (\ref{eq:A_PIPI})
and its value is given by (\ref{eq:table_PIPI_E}).
The dimensionless constant $F^{I}$
is the Lellouch-L\"uscher factor~\cite{LL-factor:LL}
given by
\begin{eqnarray}
  ( F^{I} )^2
     &=& \langle K | K \rangle \cdot
         \langle \pi\pi ; I | \pi\pi ; I \rangle / V^2
\cr
     &=& (4\pi)\left( \frac{ ( E^{I}_{\pi\pi} )^2 \, m_K }{ p^3 } \right)
               \left( p \frac{ \partial\delta^{I}(p) }{\partial p}
                    + q \frac{ \partial\phi(q) }{\partial q} \right)
 \ ,
\label{eq:F_LL}
\end{eqnarray}
where $V$ is the lattice volume $V=L^3$,
$\delta^{I}(p)$ is the two-pion scattering phase shift
for the two-pion system with the isospin $I$
at the scattering momentum $p^2=(E_{\pi\pi}^I )^2/4 - m_\pi^2$,
and $\phi(q)$ is the function defined by
\begin{equation}
  \tan\phi(q) = - \pi^{3/2} q / Z_{00}(1;q)
\ ,
\end{equation}
with the spherical zeta function,
\begin{equation}
  Z_{00}(s;q) = \frac{1}{\sqrt{4\pi}} \sum_{{\bf n}\in \mathbb{Z}^3 } ( n^2 - q^2 )^{-s}
\ ,
\end{equation}
at $q=p (2\pi/L)$.
In the noninteracting two-pion case,
the factor takes the form
$(F^{I})^2 \equiv ( F|_{\rm free} )^2 = (2 m_K V )\cdot ( 2 m_\pi V )^2 / V^2$.

For the $I=0$ channel the statistics in the present work
is not sufficient to obtain the scattering phase shift.
We therefore use the factor for the noninteracting case,
leaving a precise estimation of the factor to study in the future.
For the $I=2$ case
the phase shift is obtained with a sufficient statistics at the needed momentum.
Because the scattering momentum $p$ takes a small value,
$p=2.053(97)\times 10^{-2}$ ($44.7(2.1)\,{\rm MeV}$) in our case,
the phase shift can be approximated by
$\delta^{I=2}(p) = p (\partial \delta^{I=2}(p) / \partial p) + {\rm O}(p^3)$.
We neglect the cubic term, and find $F^{I=2} / F|_{\rm free} = 0.9254(62)$.

\begin{figure}[h]
\includegraphics[height=6.0cm]{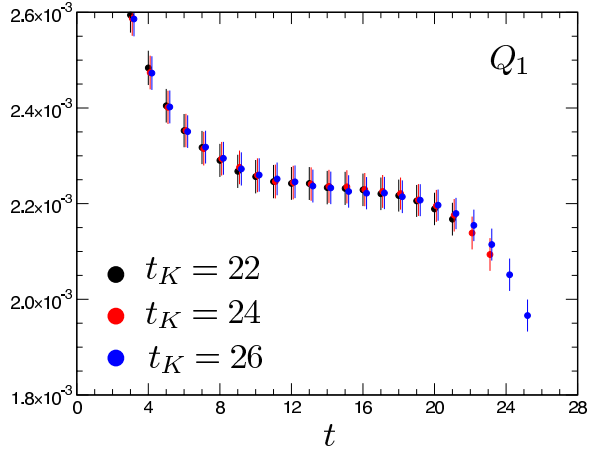}
\caption{
Effective matrix element
of $M_{1}^{I=2}(t)$ in (\ref{eq:effective_amp}),
in the lattice unit.
The time slice of the two-pion is set at $t_\pi=0$.
The operator $Q_i(t)$ runs over the whole time extent.
The matrix elements given with the $K$ meson at time slice $t_K=22,24,26$ are plotted.
}
\label{fig:AMP_1_I2}
\end{figure}

\begin{figure}[h]
\includegraphics[height=6.0cm]{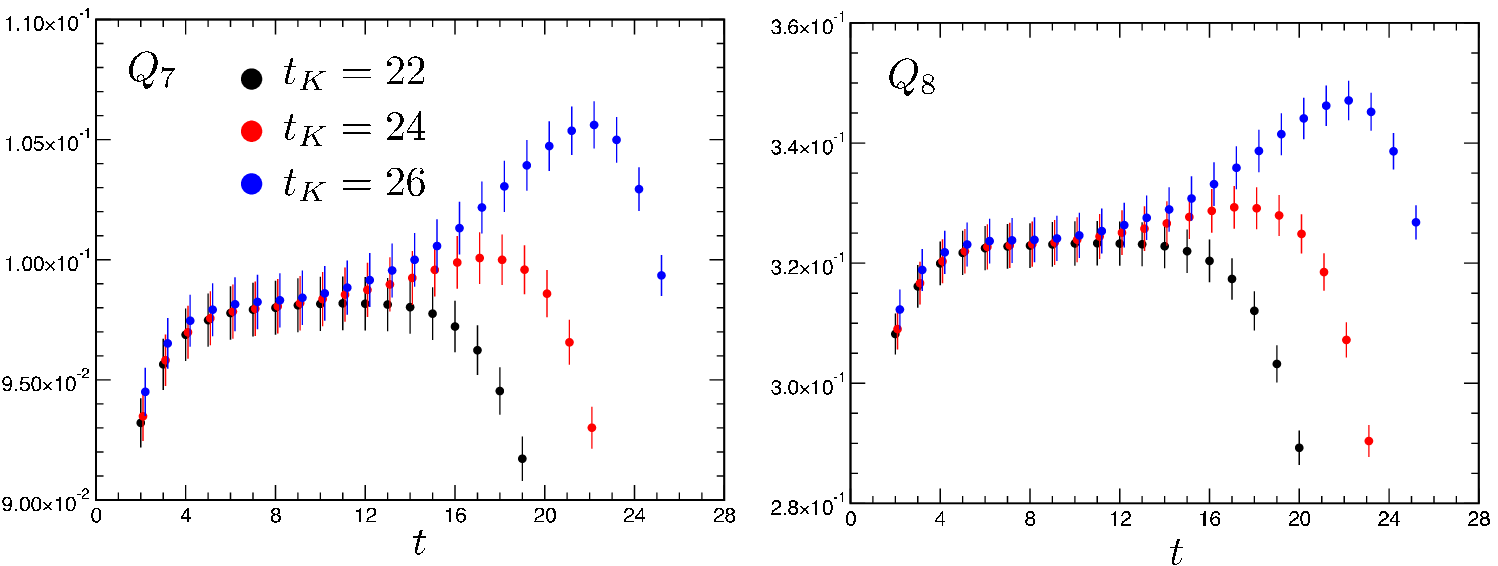}
\caption{
Effective matrix elements
of $M_{7,8}^{I=2}(t)$ in (\ref{eq:effective_amp}),
following the same convention as in Fig.~\ref{fig:AMP_1_I2}.
}
\label{fig:AMP_78_I2}
\end{figure}

\begin{figure}[h]
\includegraphics[height=6.0cm]{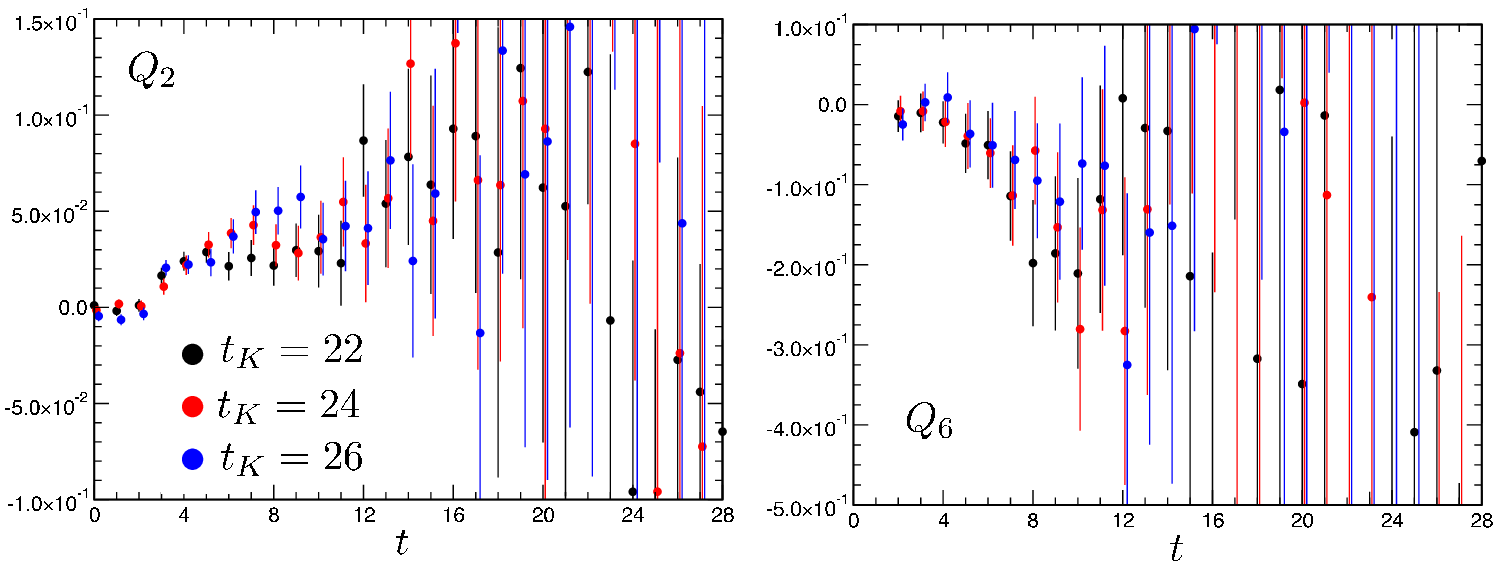}
\caption{
Effective matrix elements
of $M_{2,6}^{I=0}(t)$ in (\ref{eq:effective_amp}),
following the same convention as in Fig.~\protect\ref{fig:AMP_1_I2}.
}
\label{fig:AMP_26_I0}
%
\newpage
\end{figure}

Our results for the effective matrix elements for several representative channels
are shown in
Fig.~\ref{fig:AMP_1_I2} and Fig.~\ref{fig:AMP_78_I2} for $I=2$, and
in Fig.~\ref{fig:AMP_26_I0} for $I=0$,
where the matrix elements calculated with $t_K=22$, $24$, and $26$
are plotted.
We find plateaux for the effective matrix elements
over the time interval $t=[9,12]$ which are independent of the value of $t_K$.
This means that the around-the-world effect of the two-pion operator
is negligible in this time region.

We extract the matrix element
$M^I_i \equiv \langle K |\, \overline{Q}_i({\bf 0},0)\, | \pi\pi; I \rangle$
by a constant fit of the effective amplitude for $t_K=24$
in the time interval $t=[9,12]$.
Our results for the $I=2$ channel (the $\Delta I=3/2$ process) are
tabulated in the second column in Table~\ref{table:Result_A2},
where the relation among the matrix elements (\ref{eq:cont_KPIPI_I2_Q12}) is used.
The results for the $I=0$ channel (the $\Delta I=1/2$ process) are
tabulated in the second column in Table~\ref{table:Result_A0}.
Here we do not use the operator relations
(\ref{eq:op_rel1})--(\ref{eq:op_rel3}),
and treat each of $10$ operators as independent.
%
\subsection{ $K\to \pi\pi$ decay amplitudes }
\label{Sec: K to pipi decay amplitudes}
The renormalized matrix elements
$\overline{M}_{i}^{\,\,I}(\mu)$ are obtained
from the bare matrix elements on the lattice
$M_j^I$ extracted in the previous section
by multiplying with the renormalization factors as
\begin{equation}
  \overline{M}_{i}^{I}(q^*) =\sum_{j=1}^{10} \, M_j^{I} \, Z_{ji}(q^* a)
\ .
\label{eq:renomal_M}
\end{equation}
The renormalization factors $Z_{ij}(q^* a)$
for our choice of the fermion and gluon actions
have been calculated
by perturbation theory in one-loop order
in Ref.~\cite{Z-fact:Taniguchi}.
A nonperturbatively determination is not yet available.
For the renormalization in the continuum theory,
we adopt
the modified minimal subtraction scheme ($\overline{\rm MS}$)
with naive dimensional regularization scheme (NDR).
We choose two values $q^*=1/a$ and $\pi/a$ as the matching scale
from the lattice to the continuum theory in order to
estimate the systematic error coming
from higher orders of perturbation theory.
Large tadpole contributions
in the renormalization factors for the lattice perturbation theory are subtracted
by the mean-field improvement.
We use a mean-field improved value
in the $\overline{\rm MS}$ scheme for the coupling constant, which is given
from the bare coupling constant $g^2$
by
\begin{equation}
  1/g_{\overline{\rm MS}}^2(q^*) =
    ( C_0 P + 8C_1 R )/g^2 - 0.1006 + 0.03149 \cdot N_f
    + ( 11 - 2N_f/3 )/(8\pi^2) \cdot \log(q^* a)
\end{equation}
for our gluon and fermion actions,
where
$C_0=1-8C_1$ and $C_1=-0.331$ are the  parameters in the gluon action, and
$P$ is the expectation value of the plaquette
and $R$ is that of the $1\times 2$ Wilson loop.
The detail of the procedure was discussed in Ref.~\cite{q2-in-Z-fact:PACS-CS}.
From the values $P=0.572059(31)$ and $R=0.338902(47)$
given in Ref.~\cite{conf:PACS-CS},
we obtain
$g_{\overline{\rm MS}}^2=2.699$ at $q^*=  1/a$ and
$g_{\overline{\rm MS}}^2=1.996$ at $q^*=\pi/a$.

The decay amplitudes $A_I$ ($I=0,2$) are calculated
from (\ref{eq:weak_Hamlitonian}) as
\begin{equation}
  A_I = \sum_{i,j=1}^{10} \, \overline{M}_{i}^{\,\,I}(q^*) \, U_{ij}(q^*, \mu) C_j(\mu)
\ ,
\label{eq:A_I_from_Mbar}
\end{equation}
where
\begin{equation}
   C_i(\mu) = \frac{G_F}{\sqrt{2}}
   \left( V_{us}^* V_{ud} \right)
   \left( z_i(\mu) + \tau y_{i}(\mu) \right)
\ .
\label{eq:def_C}
\end{equation}
The explicit form of the functions $z_i(\mu)$ and $y_i(\mu)$
in the NDR scheme have been given in Ref.~\cite{Review:Buras}.
The functions $U_{ij}(q^*,\mu)$ are the running factor
of the operators $Q_i$ from the scale $q^*$ to $\mu$
for the number of  active quark flavors equal to $N_f=3$,
which have also been given in Ref.~\cite{Review:Buras}.
In the present work
we set $\mu=m_c=1.3\,{\rm GeV}$ in (\ref{eq:A_I_from_Mbar})
and evaluate the two functions $z_i(\mu)$ and $y_i(\mu)$
with the Standard Model parameters tabulated in Table~\ref{table:STD-param}.
We adopt the standard representation of the CKM matrix,
in which $CP$ violation enters entirely through the complex phase of
$V_{td}$, thus
$\tau = - \left( V_{ts}^* V_{td} \right) / \left( V_{us}^* V_{ud} \right)$.
The values of the two functions are tabulated in Table~\ref{table:coef_zy}.

From (\ref{eq:renomal_M}) and (\ref{eq:A_I_from_Mbar})
the decay amplitudes can be written in terms of  the bare matrix element $M_i^I$ as
\begin{eqnarray}
   A_I = \sum_{i=1}^{10} \, M_i^I\, \overline{C}_{i} = \sum_{i=1}^{10} A_I(i)
  \quad , \quad ( \,\, A_I(i) = M_i^I\, \overline{C}_{i} \,\, )
\ ,
\label{eq:A_I_from_M}
\end{eqnarray}
where
\begin{equation}
  \overline{C}_{i} = \sum_{j,k=1}^{10}\, Z_{ij}(q^* a) \, U_{jk}(q^*,\mu)\, C_k(\mu)
\ .
\label{eq:def_Cbar}
\end{equation}
The constant $\overline{C}_i$ should be independent of $\mu$ and $q^*$,
and depend only on the lattice cutoff $1/a$,
if we work in the full order of perturbation theory.
We define $\overline{z}_i$ and $\overline{y}_i$
by (\ref{eq:def_C}) for $\overline{C}_i$.
The values of these quantities for $q^*=1/a$ and $\pi/a$
at $\mu=m_c=1.3\,{\rm GeV}$ are given
in Table~\ref{table:coef_zy}.

Our final results of the decay amplitudes are given
in Table~\ref{table:final_result}.
The direct $CP$ violation parameter $\epsilon'/\epsilon$
is obtained by
\begin{equation}
   {\rm Re}(\epsilon'/\epsilon)
 = \frac{\omega}{\sqrt{2} |\epsilon| }
   \left(
       \frac{ {\rm Im}A_2 }{ {\rm Re}A_2 }
     - \frac{ {\rm Im}A_0 }{ {\rm Re}A_0 }
   \right)
\ ,
\end{equation}
with $\omega = {\rm Re}A_2/{\rm Re}A_0$,
where the experimental value of the indirect $CP$ violation parameter
$|\epsilon| = 2.228\times 10^{-3}$ is used in the estimation.
The statistical errors are estimated
by the jackknife procedure with a bin size of $10$ configurations
($250$ MD time units).
We also list
results of the RBC-UKQCD Collaboration
at the similar quark masses
( $m_\pi=422\,{\rm MeV}$~\cite{A0:RBC-UKQCD_400} and
  $      330\,{\rm MeV}$~\cite{A0:RBC-UKQCD_300} )
for comparison.
These two cases are calculated
with the unphysical kinematics at $m_K \sim 2 m_\pi$,
as in our calculation.
In the table, 
the results of the RBC-UKQCD Collaboration
for the $\Delta I=3/2$ process
obtained at the physical quark mass
with the physical kinematics,
where the pions in the final state have finite momenta,
in the continuum limit presented in Ref.~\cite{A2:RBC-UKQCD_phys_cont},
and the experimental values
are also tabulated.
We note that our results with an unphysical kinematics
can not be directly compared
with these values at the physical quark mass.

From Table~\ref{table:final_result}
we learn that the dependence on $q^*$
is negligible for most of the decay amplitudes,
but it is very large for ${\rm Im}A_2$.
A nonperturbative determination of the renormalization factor
is necessary to obtain a reliable result for this value.
We find a large enhancement of the $\Delta I=1/2$ process
over that for the $\Delta I=3/2$
at our quark mass $m_\pi = 280\,{\rm MeV}$.
The RBC-UKQCD Collaboration found that
the enhancement was explained
by the following numerical mechanism~\cite{DeltaIh_under}:
A large cancellation between two dominant quark diagrams occurs
for the $\Delta I=3/2$ process, rendering ${\rm Re}A_2$ small,
while such a cancellation does take place for the $\Delta I=1/2$ process.
We confirm this numerical mechanism also in our case.

Our result for $A_0$, particularly for the imaginary part,
still has a large statistical error
so that we do not obtain a nonzero result
for ${\rm Re}(\epsilon'/\epsilon)$ over the error.
We observe that the results for $A_0$ by the RBC-UKQCD Collaboration
at a similar pion mass $m_\pi=330\,{\rm MeV}$~\cite{A0:RBC-UKQCD_300}
have smaller errors than ours.
This is because they use a different two-pion operator for which the wall sources
for the two pions are separated by $\delta=4$ in the time direction,
and they set the fitting range closer to the two-pion source than our case
in extracting the matrix elements from the time correlation functions.
Improving statistics by devising a more efficient operator
for the two-pion state is an important work reserved for the future.

The contributions of the bare matrix element $M_i^{I}$ to
the decay amplitude $A_I$
($A_I(i)$ in (\ref{eq:A_I_from_M}))
are tabulated
in Table~\ref{table:Result_A2} for the $\Delta I=3/2$ and
in Table~\ref{table:Result_A0} for the $\Delta I=1/2$ process.
We find
that the main contribution to ${\rm Re}A_2$ comes from the operator $Q_1$ and $Q_2$,
and that to ${\rm Im}A_2$ from $Q_8$.
The main contribution to ${\rm Re}A_0$ comes from the operator $Q_2$
and that to ${\rm Im}A_0$ from $Q_6$.
%
%
\section{ Conclusions }
\label{Sec:Conclusions}
In the present work
we have shown that
mixings with four-fermion operators with wrong chirality
are absent even for the Wilson fermion action
for the parity odd process due to CPS symmetry.
Therefore, after subtraction of an effect from the lower dimensional operator,
a calculation of the decay amplitudes is possible
without additional calculations for the operator with wrong chirality.
This is the same situation
for chirally symmetric lattice actions such as the domain wall action.
A potential advantage with the Wilson fermion action
over chirally symmetric lattice actions
is that the computational cost is generally smaller.
Hence, with the same amount of computational resources,
a statistical improvement may be expected.

As the first step of a study to verify the possibility of calculations,
we considered the $K$ meson decay amplitude
for both the $\Delta I=1/2$ and $3/2$ channels
with the Wilson fermion action
at an unphysical quark mass $m_K \sim 2 m_\pi$.
We have found that the stochastic method
with the hopping parameter expansion technique
and the truncated solver method are very efficient for variance reduction,
yielding a first result for the $I=0$ amplitude with the Wilson fermion action.

We have been able to show a large enhancement of the $\Delta I=1/2$ process
over that for the $\Delta I=3/2$ at our quark mass
($m_\pi=275.7(1.5)\,{\rm MeV}$ and $m_K=579.7(1.3)\,{\rm MeV}$).
However, our result for $A_0$, particularly for the imaginary part,
still has a large statistical error
so that we have not obtained a nonzero result
for ${\rm Re}(\epsilon'/\epsilon)$ over the error.
For the $I=0$ two-pion system, 
the statistics in the present work
are not sufficient to obtain the scattering phase shift.
We therefore used the Lellouch-L\"uscher factor for the noninteracting case
in the calculation of the $\Delta I=1/2$ process.
Improving statistics by devising a more efficient operator
for the $I=0$ two-pion state is an important work reserved for the future.

Our calculation is carried out away from the physical quark masses,
and the decay of the $K$ meson to two zero momentum pions
at $m_K \sim 2 m_\pi$ is considered.
Clearly, we need to work toward smaller quark masses
and a more realistic kinematics
in which the two pions carry finite momenta.
This will be a major challenge that we now have to face.
%
%
\section*{Acknowledgments}
We would like to thank members of the RBC-UKQCD Collaboration
for kind comments and suggestions.
This research used computational resources of the K computer
provided by the RIKEN Advanced Institute for Computational Science
and T2K-TSUKUBA by University of Tsukuba
through the HPCI System Research Project (Project ID:hp120153).
This work is supported
by Grants-in-Aid of the Ministry of Education No.~23340054,
Interdisciplinary Computational Science Program in CCS, University of Tsukuba,
and
Large Scale Simulation Program No.~12-08 (FY2012)
of High Energy Accelerator Research Organization (KEK).
%
%

%
%
\newpage
%
%
%
%
%
%
%
%
%
%
%
%
%
%
%
%
%
%
%
%
%
%
%
%
%
%
\begin{table}[h]
\vspace{-15mm}
\caption{
Decay amplitude for the $\Delta I=3/2$ process.
The second column gives
the bare matrix elements $M_i^{I=2}$
for $Q_i$ in the lattice unit.
The other columns are their contribution to
$A_2$ ( $A_2(i)$ in (\ref{eq:A_I_from_M}) )
for $q^*=1/a$ and $\pi/a$.
}
\label{table:Result_A2}
%
\begin{tabular}{r r rr rr }
\hline\hline
     &
     & \multicolumn{2}{c}{$q^*=  1/a$}
     & \multicolumn{2}{c}{$q^*=\pi/a$}   \\
$ i$ & $M_i^{I}$
     & ${\rm Re}{A_2}\,({\rm GeV})$ & ${\rm Ime}{A_2}\,({\rm GeV})$
     & ${\rm Re}{A_2}\,({\rm GeV})$ & ${\rm Ime}{A_2}\,({\rm GeV})$ \\
\hline
$ 1$  &$   2.256(35)\times 10^{-3}$ &$-1.887(29)\times10^{-08}$&$ 0                      $&$-1.452(23)\times10^{-08}$&$ 0                      $ \\
$ 2$  &$ = M_{1}^{I=2}            $ &$ 4.330(68)\times10^{-08}$&$ 0                      $&$ 3.920(61)\times10^{-08}$&$ 0                      $ \\
$ 7$  &$   9.85(11) \times 10^{-2}$ &$ 1.053(12)\times10^{-10}$&$ 2.772(32)\times10^{-13}$&$ 3.172(36)\times10^{-10}$&$ 2.100(24)\times10^{-13}$ \\
$ 8$  &$   3.242(37)\times 10^{-1}$ &$-2.722(31)\times10^{-10}$&$-1.670(19)\times10^{-12}$&$-4.124(47)\times10^{-10}$&$-1.156(13)\times10^{-12}$ \\
$ 9$  &$ = 3/2 \cdot M_{1}^{I=2}  $ &$-1.140(18)\times10^{-12}$&$ 3.762(59)\times10^{-13}$&$ 3.739(58)\times10^{-12}$&$ 3.409(53)\times10^{-13}$ \\
$10$  &$ = 3/2 \cdot M_{1}^{I=2}  $ &$ 3.771(59)\times10^{-10}$&$-1.756(27)\times10^{-13}$&$ 4.372(68)\times10^{-10}$&$-1.409(22)\times10^{-13}$ \\
\hline
Total & \multicolumn{1}{c}{-}       &$ 2.426(38)\times10^{-08}$&$-1.192(14)\times10^{-12}$&$ 2.460(38)\times10^{-08}$&$-7.457(83)\times10^{-13}$ \\
\hline\hline
\end{tabular}
%
\end{table}
\begin{table}[h]
\caption{
Decay amplitude for the $\Delta I=1/2$ process.
The second column gives
the bare matrix elements $M_i^{I=0}$
for $Q_i$ in the lattice unit.
The other columns are their contribution to
$A_0$ ( $A_0(i)$ in (\ref{eq:A_I_from_M}) )
for $q^*=1/a$ and $\pi/a$.
}
\label{table:Result_A0}
%
\begin{tabular}{r r rr rr }
\hline\hline
     &
     & \multicolumn{2}{c}{$q^*=  1/a$}
     & \multicolumn{2}{c}{$q^*=\pi/a$}   \\
$ i$ & $M_i^{I}$
     & ${\rm Re}{A_0}\,({\rm GeV})$ & ${\rm Ime}{A_0}\,({\rm GeV})$
     & ${\rm Re}{A_0}\,({\rm GeV})$ & ${\rm Ime}{A_0}\,({\rm GeV})$ \\
\hline
$ 1$&$ 0.5(1.3)\times10^{-2}$&$-0.4(1.1)\times10^{-07}$&$  0                    $&$-3.1(8.5)\times10^{-08}$&$   0                   $\\
$ 2$&$ 3.6(1.4)\times10^{-2}$&$ 6.8(2.8)\times10^{-07}$&$  0                    $&$ 6.2(2.5)\times10^{-07}$&$   0                   $\\
$ 3$&$ 7.2(3.7)\times10^{-2}$&$-1.25(65)\times10^{-08}$&$-2.5(1.3)\times10^{-11}$&$-1.7(8.7)\times10^{-08}$&$-2.1(1.1)\times10^{-11}$\\
$ 4$&$ 1.06(40)\times10^{-1}$&$ 5.3(2.0)\times10^{-08}$&$ 6.6(2.5)\times10^{-11}$&$ 6.2(2.4)\times10^{-08}$&$ 6.1(2.3)\times10^{-11}$\\
$ 5$&$-1.0(4.3)\times10^{-2}$&$ 1.5(5.9)\times10^{-09}$&$ 1.7(6.8)\times10^{-12}$&$ 1.9(7.4)\times10^{-09}$&$ 1.8(7.1)\times10^{-12}$\\
$ 6$&$-2.0(1.1)\times10^{-1}$&$-8.4(4.6)\times10^{-08}$&$-1.03(56)\times10^{-10}$&$-7.7(4.2)\times10^{-08}$&$-8.8(4.8)\times10^{-11}$\\
$ 7$&$ 2.42(18)\times10^{-1}$&$ 2.58(19)\times10^{-10}$&$ 6.81(50)\times10^{-13}$&$ 7.79(57)\times10^{-10}$&$ 5.16(38)\times10^{-13}$\\
$ 8$&$ 7.46(54)\times10^{-1}$&$-6.26(45)\times10^{-10}$&$-3.84(28)\times10^{-12}$&$-9.48(68)\times10^{-10}$&$-2.66(19)\times10^{-12}$\\
$ 9$&$-3.0(1.4)\times10^{-2}$&$ 1.02(48)\times10^{-11}$&$-3.4(1.6)\times10^{-12}$&$-3.4(1.6)\times10^{-11}$&$-3.1(1.4)\times10^{-12}$\\
$10$&$ 0.0(1.2)\times10^{-2}$&$ 0.0(1.4)\times10^{-11}$&$-0.1(6.4)\times10^{-13}$&$ 0.0(1.6)\times10^{-11}$&$-0.1(5.2)\times10^{-13}$\\
\hline
Total & \multicolumn{1}{c}{-} &$ 6.0(3.6)\times10^{-07}$&$ -6.7(5.6)\times10^{-11}$&$ 5.6(3.2)\times10^{-07}$&$ -5.2(4.8)\times10^{-11}$\\
\hline\hline
\end{tabular}
%
\newpage
\end{table}
\begin{table}[h]
\caption{
Standard model parameters
used to evaluate the decay amplitudes
in the present work (from Ref.~\cite{STD-param:PDG}).
$\tau = - \left( V_{ts}^* V_{td} \right) / \left( V_{us}^* V_{ud} \right)$
and $\Lambda^{(5)}_{\overline{\rm MS}}$ is the lambda QCD for $N_f=5$ theory.
The standard representation of the CKM matrix of Ref.~\cite{STD-param:PDG}
is adopted,
where the $CP$ violation enters entirely through a complex phase of
$V_{td}$, thus $\tau$.
}
\label{table:STD-param}
%
\begin{tabular}{lr}
\hline\hline
$ m_Z $  &  $  91.188 \,{\rm GeV} $ \\
$ m_W $  &  $  80.385 \,{\rm GeV} $ \\
$ m_t $  &  $   173   \,{\rm GeV} $ \\
$ m_b $  &  $   4.2   \,{\rm GeV} $ \\
$ m_c $  &  $   1.3   \,{\rm GeV} $ \\
$ \Lambda^{(5)}_{\overline{\rm MS}} $ &  $ 0.23135 \,{\rm GeV}$ \\
$ \alpha $ ( at $\mu=m_W$ )    &  $ 1 / 129 $ \\
$ \sin^2\theta_W $  &  $ 0.230     $    \\
$ G_F            $  &  $ 1.166 \times 10^{-5} \,{\rm GeV}^{-2} $ \\
$ V_{ud}         $  &  $  0.97427  $ \\
$ V_{us}         $  &  $  0.22534  $ \\
$ {\rm Re}(\tau) $  &  $  0.001513 $ \\
$ {\rm Im}(\tau) $  &  $ -0.000601 $ \\
\hline\hline
\end{tabular}
%
\end{table}
%
\begin{table}[h]
\caption{
$z_i(\mu)$, $y_i(\mu)$,
$\overline{z}_i$ and $\overline{y}_i$.
The parameters of the standard model
tabulated in Table~\ref{table:STD-param} are used in the calculations.
We set $\mu=m_c=1.3\,{\rm GeV}$, and choose two values $q^*=1/a$ and $\pi/a$
as the matching scale from the lattice to the continuum.
}
\label{table:coef_zy}
%
\begin{tabular}{ r rr rr rr }
\hline\hline
     &            &               & \multicolumn{2}{c}{$q^*=  1/a$}
                                  & \multicolumn{2}{c}{$q^*=\pi/a$}   \\
$ i$ & $z_i(\mu)$ & $y_i(\mu)$    & $\overline{z}_i$ & $\overline{y}_i$
                                  & $\overline{z}_i$ & $\overline{y}_i$  \\
\hline
$ 1$&$-4.184\times10^{-1}$&$ 0                 $&$-4.487\times10^{-1}$&$ 0                 $&$-3.453\times10^{-1}$&$ 0                  $\\
$ 2$&$ 1.218\times10^{+0}$&$ 0                 $&$ 1.029\times10^{+0}$&$ 0                 $&$ 9.321\times10^{-1}$&$ 0                 $\\
$ 3$&$ 4.575\times10^{-3}$&$ 2.910\times10^{-2}$&$-9.327\times10^{-3}$&$ 3.145\times10^{-2}$&$-1.246\times10^{-2}$&$ 2.629\times10^{-2}$\\
$ 4$&$-1.373\times10^{-2}$&$-5.782\times10^{-2}$&$ 2.703\times10^{-2}$&$-5.628\times10^{-2}$&$ 3.173\times10^{-2}$&$-5.108\times10^{-2}$\\
$ 5$&$ 4.575\times10^{-3}$&$ 4.869\times10^{-3}$&$-7.309\times10^{-3}$&$ 1.402\times10^{-2}$&$-9.284\times10^{-3}$&$ 1.470\times10^{-2}$\\
$ 6$&$-1.373\times10^{-2}$&$-9.009\times10^{-2}$&$ 2.323\times10^{-2}$&$-4.700\times10^{-2}$&$ 2.130\times10^{-2}$&$-4.015\times10^{-2}$\\
$ 7$&$ 6.305\times10^{-5}$&$-2.010\times10^{-4}$&$ 5.777\times10^{-5}$&$-2.514\times10^{-4}$&$ 1.731\times10^{-4}$&$-1.904\times10^{-4}$\\
$ 8$&$ 0                 $&$ 1.098\times10^{-3}$&$-4.573\times10^{-5}$&$ 4.600\times10^{-4}$&$-6.871\times10^{-5}$&$ 3.183\times10^{-4}$\\
$ 9$&$ 6.305\times10^{-5}$&$-1.168\times10^{-2}$&$-3.047\times10^{-6}$&$-9.925\times10^{-3}$&$ 7.287\times10^{-5}$&$-8.995\times10^{-3}$\\
$10$&$ 0                 $&$ 4.357\times10^{-3}$&$ 5.277\times10^{-5}$&$ 4.635\times10^{-3}$&$ 6.368\times10^{-5}$&$ 3.717\times10^{-3}$\\
\hline\hline
\end{tabular}
%
\newpage
\end{table}
%
%
%
\begin{table}[h]
\caption{
Results of the $K\to\pi\pi$ decay amplitudes.
The results
by the RBC-UKQCD Collaboration
at $m_\pi=422\,{\rm MeV}$~\cite{A0:RBC-UKQCD_400},
   $      330\,{\rm MeV}$~\cite{A0:RBC-UKQCD_300},
the physical quark mass in the continuum limit
(only for the $\Delta I=3/2$ process)~\cite{A2:RBC-UKQCD_phys_cont},
and the experimental values are also tabulated.
}
\label{table:final_result}
%
\begin{tabular}{l rrrrrr}
\hline
\hline
& $q^* =1/a$
& $q^* =\pi/a$
& \multicolumn{3}{c}{RBC-UKQCD}
& Exp.
\\
\hline
$a\,({\rm fm})$
& \multicolumn{2}{c}{$ 0.091 $}
& $ 0.114 $
& $ 0.114 $
& -
& -
\\
$m_{\pi}\,({\rm MeV})$
& \multicolumn{2}{c}{$ 280 $}
& 330
& 422
& 140
& 140
\\
${\rm Re}A_2 \,(\times10^{-8}\,{\rm GeV})$
& $ 2.426(38) $
& $ 2.460(38) $
& $ 2.668(14) $
& $ 4.911(31) $
& $ 1.50(4)(14) $
& $ 1.479(4)  $
\\
${\rm Re}A_0 \,(\times10^{-8}\,{\rm GeV})$
& $ 60(36)    $
& $ 56(32)    $
& $ 31.1(4.5) $
& $ 38.0(8.2) $
&
& $ 33.2(2)   $
\\
${\rm Re}A_0 / {\rm Re}A_2$
& $ 25(15)    $
& $ 23(13)    $
& $ 12.0(1.7) $
& $  7.7(1.7) $
&
& $ 22.45(6)  $
\\
${\rm Im}A_2 \,(\times10^{-12}\,{\rm GeV})$
& $ -1.192(14)  $
& $ -0.7457(83) $
& $ -0.6509(34) $
& $ -0.5502(40) $
& $ -0.699(20)(84) $
&
\\
${\rm Im}A_0 \,(\times10^{-12}\,{\rm GeV})$
& $ -67(56) $
& $ -52(48) $
& $ -33(15) $
& $ -25(22) $
&
\\
${\rm Re}(\epsilon'/\epsilon)(\times 10^{-3})$
& $ 0.8(2.5) $
& $ 0.9(2.5) $
& $ 2.0(1.7) $
& $ 2.7(2.6) $
&
& $ 1.66(23) $
\\
\hline
\hline
\end{tabular}
%
\end{table}
%
%
\end{document}